\documentclass[aps,prd,preprint,superscriptaddress,showpacs,ctexart]{revtex4-1}
\pdfoutput=1 
\usepackage{float}
\usepackage{caption}
\usepackage{textcomp}
\usepackage{amsmath}
\usepackage{amssymb}
\usepackage{graphicx}
\usepackage{subfigure}
\usepackage{diagbox}
\usepackage{color}
\usepackage{ulem}
\usepackage{setspace}
\usepackage[unicode=true, bookmarks=false, breaklinks=false,pdfborder={0 0 1},colorlinks=false] {hyperref}

\makeatletter

\providecommand{\tabularnewline}{\\}

\newcommand{\bmat}{\left(\begin{array}}
\newcommand{\emat}{\end{array}\right)}
\newcommand{\beq}{\begin{equation}}
\newcommand{\eeq}{\end{equation}}

\newcommand{\dd}{\mathrm{d}}
\def\tr{\mathrm{tr}}
\def\alt{\mathrel{\mathpalette\gl@align<}}
\def\agt{\mathrel{\mathpalette\gl@align>}}
\def\gl@align#1#2{\lower.6ex\vbox{\baselineskip\z@skip\lineskip\z@
\ialign{$\m@th#1\hfil##\hfil$\crcr#2\crcr\sim\crcr}}}

\def\su5u1{SU(5) \times U(1)}
\def\fsu5u1{SU(5) \times U(1)'}
\def\so10{SO(10)}
\def\sq20{SO(10) \times SO(10)}

\def\bwt{\begin{widetext}}
\def\ewt{\end{widetext}}
\def\be{\begin{equation}}
\def\ee{\end{equation}}
\def\bea{\begin{eqnarray}}
\def\eea{\end{eqnarray}}
\def\bean{\begin{eqnarray*}}
\def\eean{\end{eqnarray*}}
\def\bary{\begin{array}}
\def\eary{\end{array}}
\def\bit{\begin{itemize}}
\def\eit{\end{itemize}}

\makeatother

\begin{document}

\title{Natural Higgs Inflation, Gauge Coupling Unification, and Neutrino masses}

\author{Heng-Yu Chen}

\affiliation{ Bartol Research Institute, Department of Physics and Astronomy,
University of Delaware, Newark, DE 19716, USA }

\author{Ilia Gogoladze}

\affiliation{ Bartol Research Institute, Department of Physics and Astronomy,
University of Delaware, Newark, DE 19716, USA }

\author{Shan Hu}

\affiliation{Department of Physics, Faculty of Physics and Electronic Sciences, Hubei University,
Wuhan 430062, P. R. China }

\author{Tianjun Li}

\affiliation{ CAS Key Laboratory of Theoretical Physics, Institute of Theoretical
Physics, Chinese Academy of Sciences, Beijing 100190, China }

\affiliation{ School of Physical Sciences, University of Chinese Academy of Sciences,
No.~19A Yuquan Road, Beijing 100049, China }

\author{Lina Wu}

\affiliation{ Department of Applied Physics, School of Physics, University of Electronic Science and Technology of China, Chengdu 610054, P. R. China }

\affiliation{CAS Key Laboratory of Theoretical Physics, Institute of Theoretical
Physics, Chinese Academy of Sciences, Beijing 100190, China }

\date{\today}
\begin{abstract}
We present a class of non-supersymmetric models in which so-called critical Higgs inflation ($\xi<100$) naturally can be realized without using 
specific values for Higgs and top quark masses.
In  these scenarios,  the Standard Model (SM)  vacuum stability
problem, gauge coupling unification, neutrino mass generation and Higgs inflation mechanism
are linked to each other.  We adopt in our models Type I seesaw mechanism for neutrino
masses. An appropriate choice of the Type I Seesaw scale allows us to
have an arbitrarily small but positive value of SM Higgs quartic coupling
around the inflation scale.  We present
a few benchmark points where we show that the scalar spectral indices
are around 0.9626 and 0.9685 for the number of e-folding
$N=50$ and $N=60$ respectively. The tensor-to-scalar ratios are order
of $10^{-3}$. The running of the scalar spectral index is negative
and is order of $10^{-4}$.
\end{abstract}
\maketitle

\section{Introduction}

Discovery of the Higgs boson as predicted by the Standard Model (SM)
by the ATLAS and the CMS experiments became the moment of triumph
for particle physics \cite{ATLAS,CMS,moriond2013}. Such a historic
discovery together with decades of electroweak precision data have
well established the validity of SM up to accessible energies. However,
there is no verified explanation of the origin of the small neutrino
masses or inflation. No viable candidate for the dark matter in the
SM. More theoretical question: Why in the SM Higgs vacua is meta-stable
for the central values Higgs and top quark masses \cite{Buttazzo:2013uya} or why gauge couplings
does not unify precisely at high scale when we have very strong trend
for it. Due to these unwavering issues, various extensions of SM have
been proposed.

Although there are many observational evidences in favor of inflation \cite{Staro},
the nature of the inflation is still unclear. An interesting proposal
directly connects cosmology and particle physics is the idea the SM
Higgs boson could play the role of the inflaton \cite{Bezrukov:2007ep}
above a scale~$\Lambda$. Above $\Lambda$, the potential of the
Higgs field becomes flat and the slow-roll inflation is realized.
In this scenario the SM Higgs boson has non-minimal coupling $\xi$
to Ricci scalar. In general non-minima $\xi$ coupling is order of
$10^{4}$ and the unitarity is violated at the scale $M_{Pl}/\sqrt{\xi}$.
It was pointed out in ref. \cite{Bezrukov:2010jz} this result does
not necessarily spoil the self-consistency of the Higgs inflationary
scenario. On the other hand it was shown that a simple extension \cite{Giudice:2010ka}
of the SM can preserve unitarity of theory up to Planck scale.

The aim of this paper is to construct the model where so-called critical
\cite{Bezrukov:2014bra} Higgs inflation ($\xi<100$) naturally can
be realized without using specific values for Higgs and top quark
masses. When all non-minimal couplings are not particularly large,
$\xi<10^{2}$, the renormalizable low-energy effective field theory
is reliable up to $3\times10^{17}\;$GeV or so, which is close by
to the typical string scale. We also consider the case when unitarity
is valid up to Planck scale without going much in detail.

The idea of grand unification theory (GUT) or string theory is one
of the most attractive idea in particle physics. In most cases, both
theory naturally predict gauge coupling unification at high scale.
In this paper we adopt this gauge coupling unification without asking
what is underline theory. There are class of models where gauge
couplings can be unified at the Planck scale or so \cite{Haba:2014oxa}.
In this paper we present two models. They differ  by low
scale vector-like fermion content. In order to have string
or Planck scale unification we also introduce at intermediate scale
additional sfermion. The presence of these fermions significantly
modifies the evolution of SM Higgs quartic coupling and solves the
SM vacuum stability problem \cite{Degrassi:2012ry}.

The secret of neutrino masses may lie in some form of seesaw mechanism.
 For instance, SM singlet right-handed neutrinos with large Majorana masses
cause the light neutrino masses (Type-I seesaw) \cite{Seesaw}. Embedding
Type I seesaw mechanism for neutrino mass in our scenario  allows us
to have the control of SM Higgs quartic coupling at inflation scale.
As we will show in this scenario there is no problem to have any small
values for the SM Higgs quartic coupling at inflation scale. As a
result we can have the minimal values for $\xi(<100)$ coupling when
central values for the top quark and SM Higgs mass are taken.

The remainder of this paper is organized as follows. In Section 2
we briefly outline the models. In Section 3 we present results of
renormalization group equations (RGE) evolution. Section 4 is dedicated
to the Higgs inflation. Numerical study of Higgs inflation in our
models are given in section 5. The tables in section 5 present
some benchmark points which summarize our results. Our conclusions
are presented in Section 6. In Appendix A, we briefly discuss the
RGEs related to our models.


\section{The Models}

We present two models where gauge coupling unification can be achieved
around the reduced Planck scale $2.43\times10^{18}$ GeV or $3\times10^{17}$
GeV which we can relate to the string scale. It is known that there
are class of models where gauge couplings can be unified at the Planck
scale or so \cite{Haba:2014oxa}. The presence of these fermions significantly
modifies the evolution of SM Higgs quartic coupling and solves the
SM vacuum stability problem. In our scenarios we embed Type I seesaw
mechanism for neutrino mass, which allow us to have the SM
Higgs quartic coupling positive but arbitrary small at unification
scale. This observation can be useful for lowering the values of non-minimal
coupling $\xi$ between Higgs field and Ricci scalar, which is a crucial
player in the SM Higgs inflation models.


\subsection{Model I}

Model I is an extension of the SM with additional vector-like fermions
$Q_{x}(3,2,1/6)+Q_{x}^{c}(\bar{3},2,-1/6)$ around the TeV scale,
as well as the $SU(3)_{C}$ and $SU(2)_{L}$ adjoint fermions $G_{x}(3,1,0)$
and $W_{x}(1,8,0)$ at the intermediate scales. Here the brackets
contain the $SU(3)_{c}\times SU(2)_{L}\times U(1)_{Y}$ quantum numbers
of the new particles. Particular interest to us is the TeV-scale vector-like
fermions which are kinetically accessible to the current and foreseeable
future collider energies. We also introduce 3 right handed neutrinos 
in order to realize Type I seesaw mechanism for neutrino masses.

\subsection{Model II}

In model II, we have the vector-like fermions $(Q_{x}+Q_{x}^{c})$
and $(D_{x}+D_{x}^{c})$ around the TeV scale, as well as $G_{x}$
and $W_{x}$ at the intermediate scales. Unification at $M_{GUT}\sim2\times10^{16}$
GeV of the three SM gauge couplings can be achieved by postulating
the existence of $(Q_{x}+Q_{x}^{c})$ and $(D_{x}+D_{x}^{c})$ of
 vector-like fermions at TeV scale or so Ref.~\cite{GCU,Chen:2017rpn}.
Model II also has three SM singlet right handed neutrinos for neutrino
mass generation.

New particles transform under
$SU(3)_{C}\times SU(2)_{L}\times U(1)_{Y}$ symmetry as follows
\begin{eqnarray*}
 &  & Q_{x}\left(3,2,\frac{1}{6}\right)+Q_{x}^{c}\left(\overline{3},2,-\frac{1}{6}\right),~~~D_{x}\left(3,1,-\frac{1}{3}\right)+D_{x}^{c}\left(\overline{3},1,\frac{1}{3}\right),~~\\
 &  & G_{x}\left(8,1,0\right)+W_{x}(1,3,0).~\,
\end{eqnarray*}
The TeV-scale vector-like fermions alter the RGE evolution and as
a result we have a positive value for Higgs quartic coupling all the way
up to the reduced Planck scale. We introduce $G_{x}$ and $W_{x}$
to achieve the unification scale higher than the traditional GUT scale
\cite{Haba:2014oxa}.

\subsection{Type-I seesaw}

This is the simplest extension of the SM for understanding the small
neutrino masses. It just requires 3 addition of SM-singlet Majorana
fermions, known as right handed neutrinos $N_{\alpha}$, to the SM
particle content. The relevant piece of the Lagrangian is given by
\begin{eqnarray*}
-{\cal L}\ =\ (Y_{\nu})_{\alpha\beta}\bar{L}_{\alpha}\widetilde{H}N_{\beta}+\frac{1}{2}(M_{N})_{\beta\gamma}\bar{N}_{\beta}^{c}N_{\gamma}+{\rm H.c.}\,,
\end{eqnarray*}
with $\widetilde{H}=i\sigma_{2}H^{*}$ and Greek letters stand for
family indices. After electroweak symmetry breaking, a Dirac neutrino
mass term $M_{D}=Y_{\nu}v$ which, together with the Majorana mass
$M_{N}$, induces the tree-level active neutrino masses by the seesaw
formula
\begin{align}
M_{\nu}\ \simeq\ -M_{D}M_{N}^{-1}M_{D}^{\sf T}\,.\label{seesaw}
\end{align}
Here $v$ stands for the Higgs vacuum expectation value (VEV) and
$v=246$ GeV. It is well known that in order to satisfy current experimental
data for neutrino masses Dirac Yukawa coupling $Y_{\nu}$ needs to
be $O(1)$ when right handed neutrino mass scale is $10^{14}$ GeV
or so. As we will show big Dirac Yukawa coupling will give a significant
contribution to the SM Higgs quartic coupling evolution. These
contributions allow us to have the SM Higgs quartic coupling positive and as small as possible
in order to have the successful Higgs inflation.


\section{Solution to the RGEs}

To analyze the evolution of the SM parameters, one needs to solve
the set of RGEs which in turn requires one to define the relevant
couplings at some energy scale. In this case, all the SM gauge couplings,
Yukawa couplings and SM Higgs quartic coupling were evaluated at
two-loop level up to the energy scale corresponding to the new particle
mass scale. At energies above the new particle mass the couplings
were evolved continuously but with the updated RGEs, see the appendix
A.

With $m_{t}=173.34{\rm ~GeV}$ and $m_{h}=125.06{\rm ~GeV}$, we present
the gauge coupling unification at the reduced Planck scale in Fig.~1(a)
and Fig.~2(a) for Models I and II, respectively. In Figure 1(a),
we present the evolution of the SM gauge (dash line) and top Yukawa
(solid line) couplings in the Model I. The vector-like
fermion $(Q_{x}+Q_{x}^{c})$ mass is set equal to 1 TeV. We choose
1 TeV scale for the new vector-like fermion because it is close to the
current experimental bound and we hope the models can be tested at
LHC. On the other hand it is well known that changing 1
TeV scale to a few TeV does not affect significantly the RGE evolution. 
While masses for the
adjoint fermions $G_{x}(8,1,0)$ ($m_{8}=10^{6}$ GeV) and $W_{x}(1,3,0)$
($M_{3}=1.13\times10^{12}$) GeV are obtained from requiring that
the gauge couplings should be unified at $2.43\times10^{18}$ GeV.
The scale for neutrino type I seesaw is taken at $10^{14}$ GeV. We
can see from figure 1(a) that neutrino Dirac Yukawa coupling does not give
noticeable contribution to the gauge and top Yukawa coupling RGE evolution but it is very important for the SM Higgs quartic coupling evolution.

The panel (b) in Figure 1 displays the evolution of SM Higgs
quartic ($\lambda$) coupling for Model I. Different lines correspond
to the different values of the SM Higgs and top quark mass at electroweak
scale. The solid line stands for central values for $M_{h}=125.09$
GeV and $m_{t}=173.34$ GeV. The dashed lines correspond to the $M_{h}=128.69$
GeV and $m_{t}=171.82$ GeV. The region between the dashed and solid
lines is the allowed parameter space of $m_{t}$ and $M_{h}$.

The evolution of the SM Higgs coupling in Figure 1(b) is mostly governed
by the interplay between the top Yukawa coupling and SM Higgs quartic
coupling, which have comparable and dominant contribution in the RGE
for SM Higgs quartic coupling. (See Eq. (\ref{A7})). The negative
sign contribution from the top Yukawa coupling makes the SM Higgs quartic
coupling smaller during the evolution. This is how the lower bound
for Higgs boson mass is obtained in the SM from the vacuum stability
condition. On the other hand, additional colored vector-like fermion
at TeV scale changes RGE evaluation slop for the SM gauge coupling
as we see from Figure 1(a). As a consequence it changes slop of top
Yukawa coupling evolution and makes it more steeper compare to the
SM case \cite{Gogoladze:2010in}. So, having in the model a smaller
value for the top Yukawa coupling in the RGE evolution means having
a milder contribution in the RGE for the SM Higgs quartic coupling.
This makes RGE evolution for the SM Higgs quartic coupling 
changing gradually and as a result the SM Higgs vacuum stability problem
can be solved. As we can see from Figure 1(a) above $10^{10}$ GeV
the value of top Yukawa becomes smaller than all SM gauge couplings,
which means gauge coupling contributions to the SM Higgs quartic coupling
RGE can become comparable. As the result what we can see in Figure
1(b) SM Higgs quartic coupling reaches a minimal value at $10^{12}$
GeV. This means at this scale all gauge, top Yukawa and SM Higgs quartic
coupling cancel each other in SM Higgs quartic coupling RGE. After
$10^{12}$ GeV, gauge coupling contributions become dominant and SM
Higgs quartic coupling starts raising. This tendency continues until
$10^{14}$ GeV which is neutrino seesaw scale in our model. In this
case, in order to have correct neutrino masses for active neutrinos,
the neutrino Dirac Yukawa coupling $Y_{\nu}$ has to be $O(1)$. On
the other hand, neutrino Dirac Yukawa couplings give contributions
to SM Higgs quartic coupling evolution. As we can see in Figure
1(b), it changes drastically the SM Higgs quartic coupling evolution.
This is why we have a strong bend there. The slop of SM Higgs quartic
coupling again has a decreasing tendency. It is clear, in our scenario,
 a suitable choice of neutrino seesaw scale allows as to have any small
but positive value for SM Higgs quartic coupling at string or reduced
Planck scale. As we will show that this observation will be very useful
when we discuss the SM Higgs inflation.

\begin{figure}[H]
\centering
\subfigure[]{\includegraphics[height=4cm]{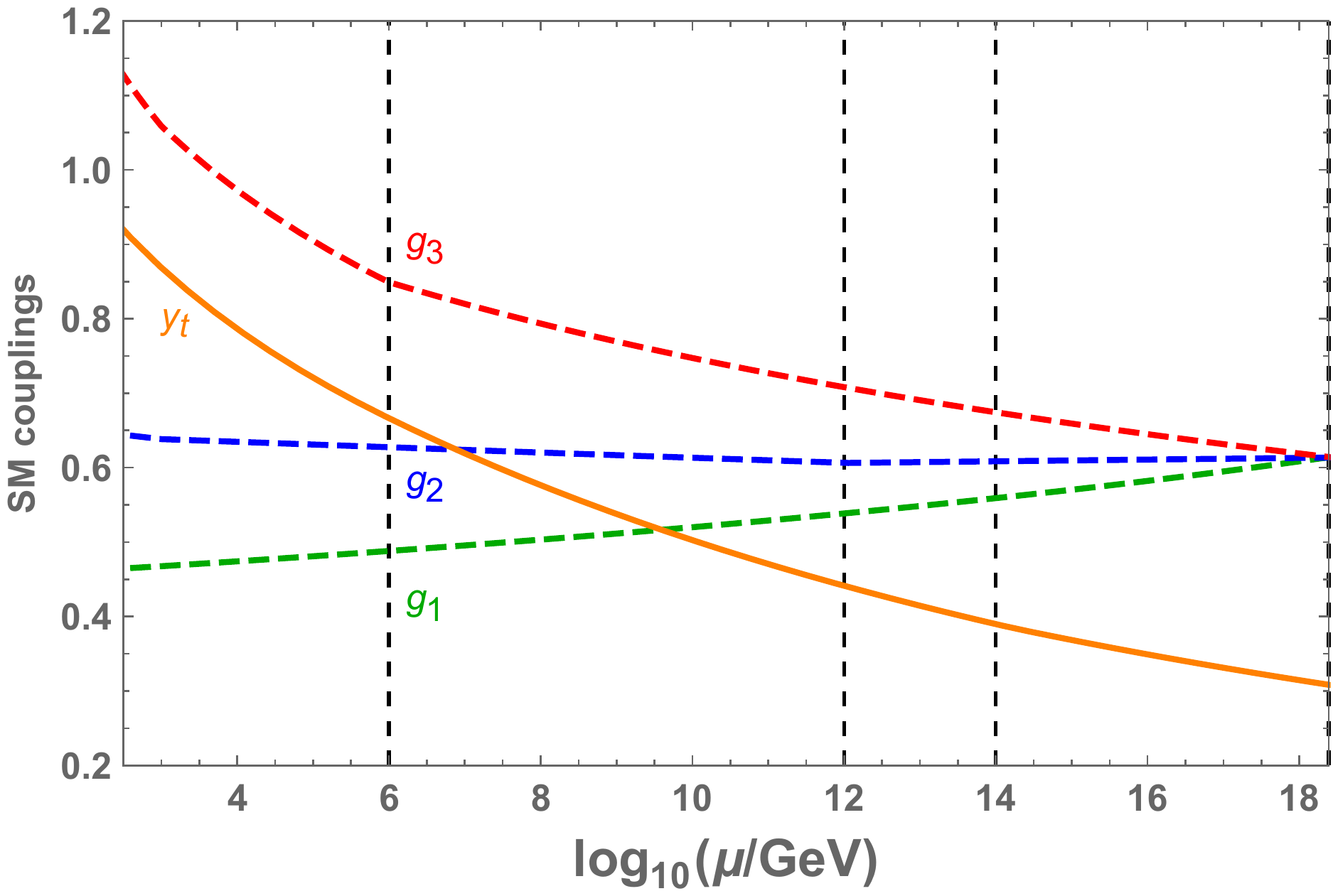}}
\subfigure[]{\includegraphics[height=4cm]{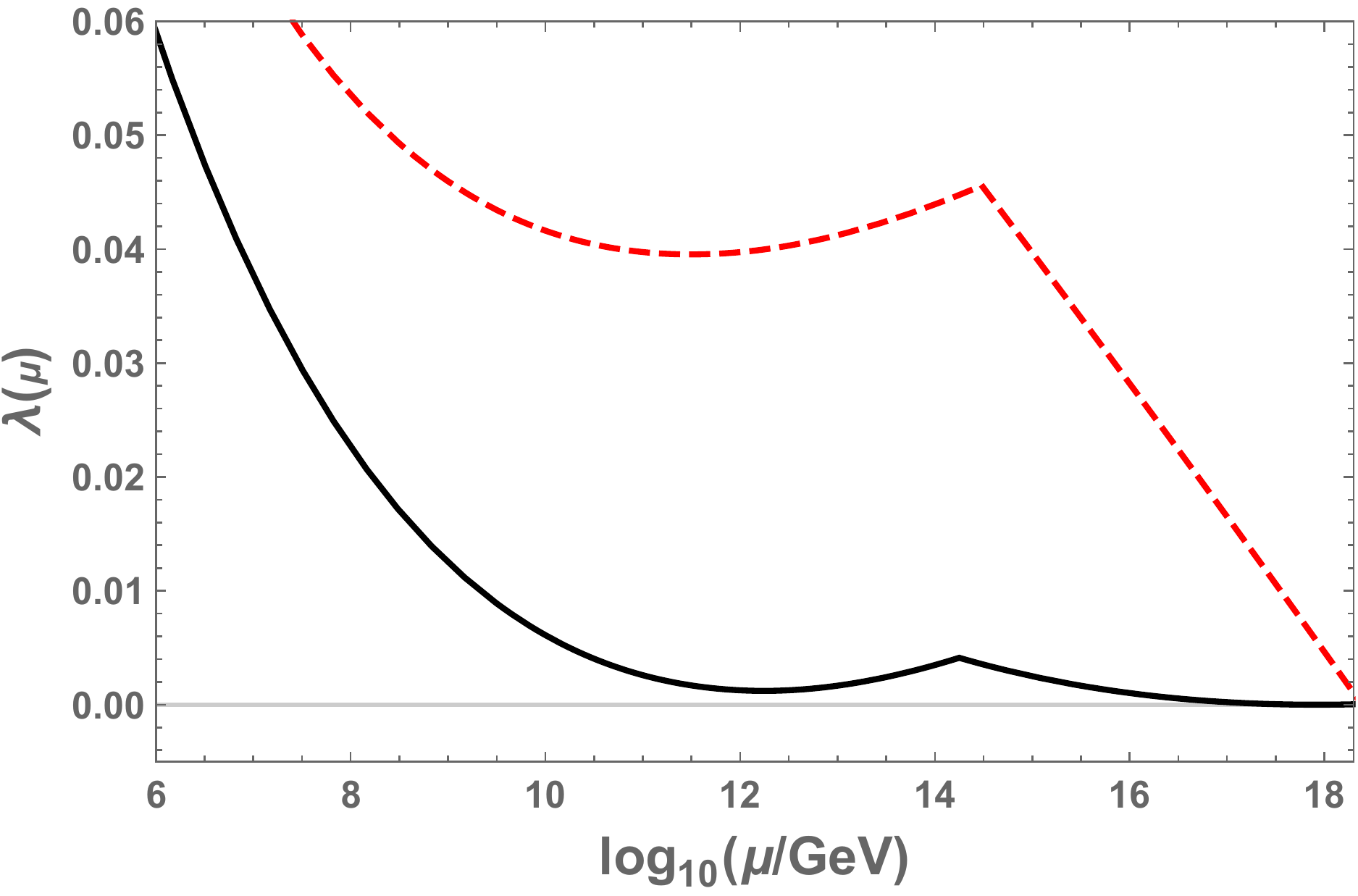}}
 \caption{The SM gauge (dash line) and top Yukawa (solid line) couplings evolution
in the {\textbf{Model I}} (panel (a)). The vector-like fermion $(Q_{x}+Q_{x}^{c})$
mass is set equal to 1 TeV. While masses for the particles $G_{x}(8,1,0)$
($m_{8}=10^{6}$ GeV) and $W_{x}(1,3,0)$ ($M_{3}=1.13\times10^{12}$)
GeV are obtained from requiring that the gauge couplings should be unified
at $2.43\times10^{18}$ GeV. The scale for neutrino type I seesaw
is taken at $10^{14}$ GeV. The panel (b) in Figure 1 displays evolution
of the SM Higgs quartic coupling for Model I. The solid line stands
for central values for $M_{h}=125.09$ GeV and $m_{t}=173.34$ GeV.
The dashed lines corresponds to the $M_{h}=128.69$ GeV and $m_{t}=171.82$
GeV. The scale of x-axis is zoomed in order to show the behavior of $\lambda$ at high energy scale.}
\label{case1} 
\end{figure}

$\;$

\begin{table}[H]
{\centering
\begin{center}\renewcommand\arraystretch{1.3}
\begin{tabular}{|c|ccccc|c|}
\hline
\diagbox{$m_{t}$}{$M_{h}$}  & 121.49  & 123.29  & 125.09  & 126.89  & 128.69  & Min.\tabularnewline
\hline
171.82  & -  & $~2.0574\times10^{14}~$  & $~2.4043\times10^{14}~$  & $~2.7232\times10^{14}~$  & $~3.0305\times10^{14}~$  & 122.02\tabularnewline
172.58  & -  & -  & $~2.1134\times10^{14}~$  & $~2.4644\times10^{14}~$  & $~2.7929\times10^{14}~$  & 123.46\tabularnewline
173.34  & -  & -  & $~1.7786\times10^{14}~$  & $~2.1759\times10^{14}~$  & $~2.5334\times10^{14}~$  & 124.89\tabularnewline
174.10  & -  & -  & -  & $\;1.8375\times10^{14}~$  & $~2.2390\times10^{14}~$  & 126.34\tabularnewline
174.86  & -  & -  & -  & -  & $~1.8977\times10^{14}~$  & 127.79\tabularnewline
\hline
\end{tabular}
 \end{center}
\caption{For \textbf{Model I}, dependence of right handed neutrino mass on the top and SM Higgs masses when gauge couplings are unified at
 $M_{U}=2.43\times10^{18}~{\rm GeV}$. We use hyphen when there is no solution for given Higgs and top quark satisfying vacuum stability condition. For all solutions presented here, the SM Higgs quartic coupling is $O(10^{-6})$ at $M_U$ scale. The last column represents  the lowest bounds of $M_{h}$ which satisfies vacuum stability bound for given top quark mass.}
\label{tab:small_xi_case1_a}}
\end{table}

In Table \ref{tab:small_xi_case1_a}, the dependence of right handed neutrino mass on the top and SM Higgs masses is shown when gauge couplings are unified at $M_{U}=2.43\times10^{18}~{\rm GeV}$. The $m_{t}$ and $M_{h}$ are taken within two sigma uncertainty. We use hyphen when there is no solution for given SM Higgs and top quark satisfying vacuum stability condition. The last column represents  the lowest bounds of $M_{h}$ which satisfies vacuum stability bound for given top quark mass. For all solutions presented in Table \ref{tab:small_xi_case1_a}, the SM Higgs quartic coupling is $O(10^{-6})$ at $M_U$ scale. This will be very important once we consider Higgs inflation in next section.

In Figure 2(a), we display results for Model II with $m_{t}=173.34{\rm ~GeV}$
and $m_{H}=125.06{\rm ~GeV}$. The SM gauge (dash line) and top Yukawa
(solid line) couplings evolution in present of vector-like fermions
$(Q_{x}+Q_{x}^{c})$ and $(D_{x}+D_{x}^{c})$ at 1 TeV. The mass scale
for $G_{x}(8,1,0)$ and $W_{x}(1,3,0)$ are the same and equal to
$2.2\times10^{12}$ GeV. In this case we have the gauge couplings unified
at $2.43\times10^{18}$ GeV. The neutrino type I seesaw scale is taken
at $1.2\times10^{14}$ GeV. The right panel (b) displays evolution
of the SM Higgs quartic coupling for Model II. Different lines correspond
to the different values of the SM Higgs and top quark mass at electroweak
scale. The solid line stands for central values for $M_{h}=125.09$
GeV and $m_{t}=173.34$ GeV. The dotted line stands for $2\sigma$
theoretical and experimental bounds \cite{Patrignani:2016xqp} $M_{h}=126.57$
GeV and $m_{t}=174.86$ GeV. The dashed lines correspond to the $M_{h}=128.69$
GeV and $m_{t}=171.82$ GeV. Behavior of RGE evolution of couplings are very similar what we had in Figure 1. Only significant difference is that additional vector-like
fermions at TeV scale make more gradual evolution of SM Higgs quartic
coupling compare to the Model I. This makes SM Higgs quartic coupling
bigger when it reaches its minimal value around $10^{11}$ GeV scale,
which allows us to have wider range for Top and Higgs masses and still
have successful SM Higgs inflation.

As we can see from Figure 2(b), the solid line is way above zero during the RGE evolution.
This makes the solution stable from radiative correction. Note the value of SM Higgs quartic coupling
with central values of top and Higgs masses in Model II is significantly larger than the value in Model I. 

\begin{figure}[H]
\centering
\subfigure[]{\includegraphics[height=4cm]{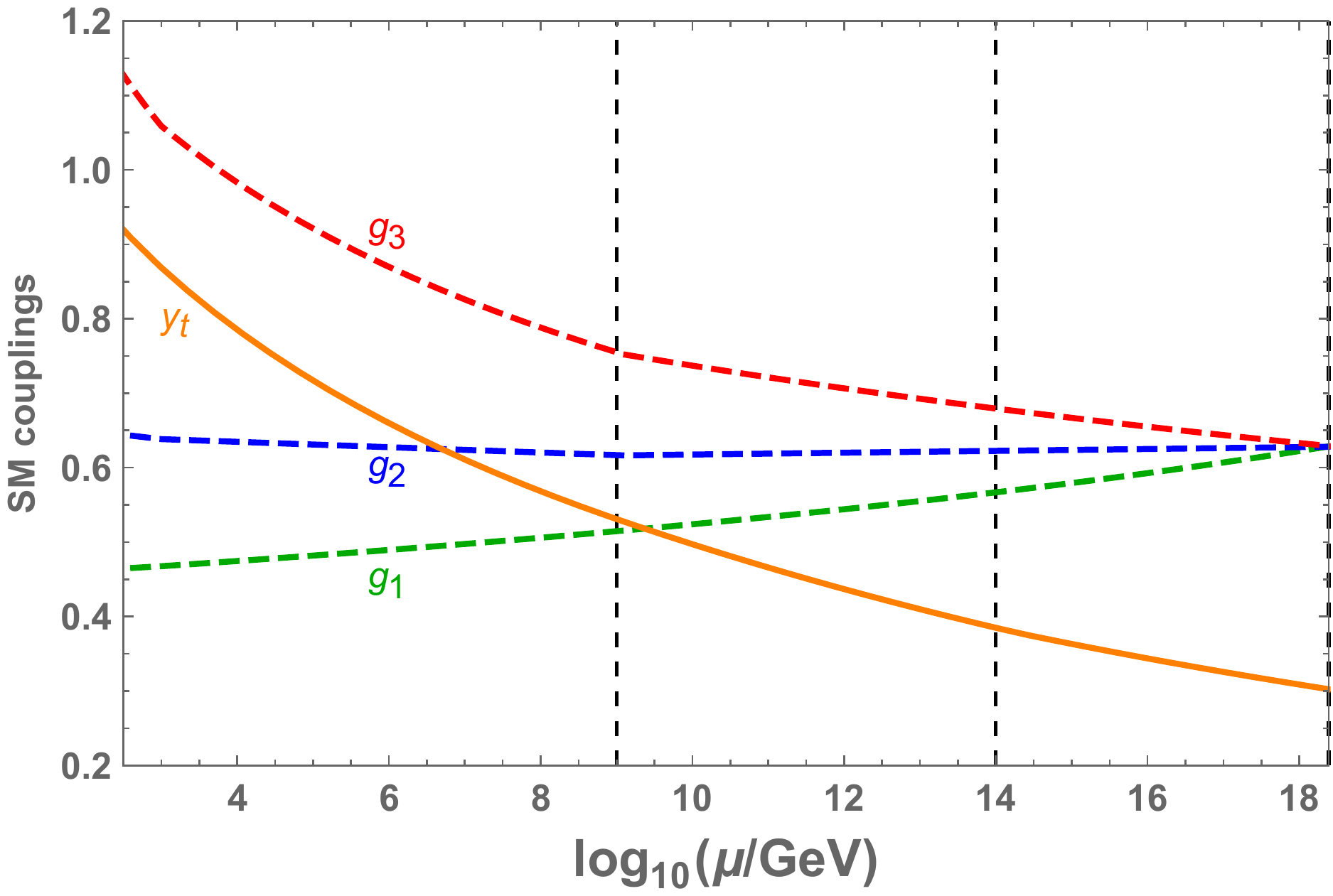}}
\subfigure[]{\includegraphics[height=4cm]{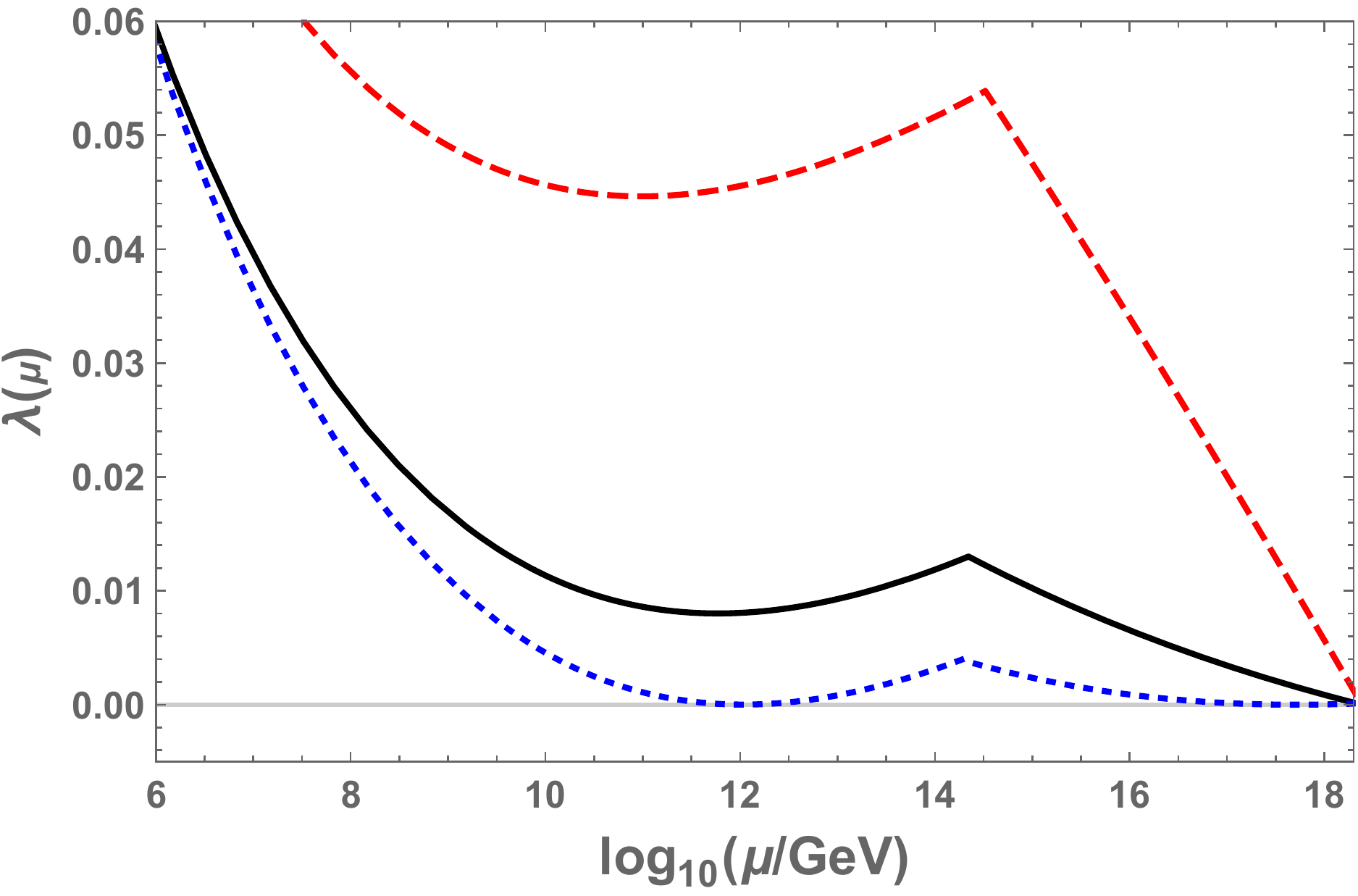}}
\caption{The SM gauge (dash line) and top Yukawa (solid line) couplings evolution
in the {\textbf{Model II}} (panel (a)). The vector-like fermion
$(Q_{x}+Q_{x}^{c})$ and $(D_{x}+D_{x}^{c})$ masses are set
to 1 TeV. The mass scale for $G_{x}(8,1,0)$ and $W_{x}(1,3,0)$ particle
masses are the same and equal to $1.2\times10^{9}$ GeV. In this case
we have the gauge couplings being unified at $2.43\times10^{18}$
GeV. The neutrino type I seesaw scale is taken at $2.2\times10^{14}$
GeV. The right panel (b) displays evolution of the SM Higgs quartic
coupling for Model II. The solid line stands for central values for
$M_{h}=125.09$ GeV and $m_{t}=173.34$ GeV. The dotted line stands
for $M_{h}=126.57$ GeV and $m_{t}=174.86$ GeV. The dashed lines
corresponds to the $M_{h}=128.69$ GeV and $m_{t}=171.82$ GeV. The scale of x-axis is zoomed in order to show the behavior of $\lambda$ at high energy scale.}
\end{figure}

\begin{table}[H]
\centering
\begin{tabular}{|c|ccccc|c|}
\hline
\diagbox{$m_{t}$}{$M_{h}$}  & 121.49  & 123.29  & 125.09  & 126.89  & 128.69  & Min.\tabularnewline
\hline
171.82  & $~2.1146\times10^{14}~$  & $~2.4309\times10^{14}~$  & $~2.7292\times10^{14}~$  & $~3.0160\times10^{14}~$  & $~3.2959\times10^{14}~$  & 121.49 \tabularnewline
172.58  & -  & $~2.1625\times10^{14}~$  & $~2.4858\times10^{14}~$  & $~2.7911\times10^{14}~$  & $~3.0851\times10^{14}~$  & 122.28\tabularnewline
173.34  & -  & -  & $~2.2135\times10^{14}~$  & $~2.5440\times10^{14}~$  & $~2.8565\times10^{14}~$  & 123.72\tabularnewline
174.10  & -  & -  & -  & $~2.2679\times10^{14}~$  & $~2.6056\times10^{14}~$  & 125.14\tabularnewline
174.86  & -  & -  & -  & $~1.9513\times10^{14}~$  & $~2.3257\times10^{14}~$  & 126.57\tabularnewline
\hline
\end{tabular}\caption{For \textbf{Model II}, dependence of right handed neutrino mass on the top and SM Higgs masses when gauge couplings are unified at  $M_{U}=2.43\times10^{18}~{\rm GeV}$. We use hyphen when there is no solution for given Higgs and top quark satisfying vacuum stability condition. For all solutions presented here, the SM Higgs quartic coupling is $O(10^{-6})$ at $M_U$ scale. The last column represents  the lowest bounds of $M_{h}$ which satisfies vacuum stability bound for given top quark mass.}
\label{tab:small_xi_case2_a}
\end{table}

{In Table \ref{tab:small_xi_case2_a}, the dependence of right handed neutrino mass on the top and SM Higgs masses is shown when gauge couplings are unified at $M_{U}=2.43\times10^{18}~{\rm GeV}$. The $m_{t}$ and $M_{h}$ are taken within two sigma uncertainty. We use hyphen when there is no solution for given Higgs and top quark satisfying vacuum stability condition. The last column represents the lowest  bounds of $M_{h}$ which satisfies vacuum stability bound for given top quark mass. For all solutions presented in Table \ref{tab:small_xi_case1_a}, the SM Higgs quartic coupling is $O(10^{-6})$ at $M_U$ scale. This will be very important once we consider Higgs inflation in next section.  Compared with Model
I, the allowed range for  $m_{t}$ and $M_{h}$ parameter space is larger in Model II.}

\section{Higgs Inflation}

{}Next, we consider the Higgs inflation by introducing a non-minimal
coupling between the Higgs doublet and gravity. In the Jordan frame,
the action is
\begin{equation}
S_{J}=\int d^{4}x\;\sqrt{-g}\left[-(\frac{1}{2}+\xi H^{+}H)R+(D_{\mu}H)^{+}(D_{\mu}H)-\frac{1}{2}\lambda(H^{+}H-\frac{v^{2}}{2})^{2}\right]~,~\,\label{S}
\end{equation}
where the reduced Planck scale $M_{{\rm Pl}}$ is set to be $1$,
and $H=(0,v+\phi)/\sqrt{2}$ is the Higgs doublet in the unitary gauge.
The physical Higgs field $\phi$ is taken to be the inflaton.

The RGE improved effective inflaton potential is
\begin{equation}
V(\phi)=\frac{\lambda(\phi)}{8}\phi^{4}
\end{equation}
with $\lambda(\phi)$ the solution of the RGE for the Higgs coupling.
In the Einstein frame with a canonical gravity sector, the theory
is described by a new inflaton field $\sigma$ that has a canonical
kinetic term. The relation between $\sigma$ and $\phi$ is
\begin{equation}
\left(\frac{\dd\sigma}{\dd\phi}\right)^{-2}=\dfrac{(1+\xi\phi^{2})^{2}}{1+(6\xi+1)\xi\phi^{2}}~.
\end{equation}
The action becomes
\begin{equation}
S_{E}=\int d^{4}x\;\sqrt{-g_{E}}\left[-\frac{1}{2}R_{E}+\frac{1}{2}(\partial\sigma)^{2}-V_{E}(\phi(\sigma))\right]\;,
\end{equation}
where
\begin{equation}
V_{E}(\phi)=\frac{\lambda(\phi)\phi^{4}}{8\left(\xi\phi^{2}+1\right)^{2}}\;.
\end{equation}
The inflationary slow-roll parameters in terms of $\phi$ are expressed
as
\begin{eqnarray}
\epsilon(\phi) & = & \frac{M_{Pl}^{2}}{2}\left(\dfrac{V'_{E}}{V_{E}\sigma'}\right),\\
\eta(\phi) & = & M_{Pl}^{2}\left[\dfrac{V''_{E}}{V_{E}(\sigma')^{2}}-\dfrac{V'_{E}\sigma''}{V_{E}(\sigma')^{3}}\right],\\
\zeta(\phi) & = & M_{Pl}^{4}\left(\dfrac{V'_{E}}{V_{E}\sigma'}\right)\left(\dfrac{V'''_{E}}{V_{E}(\sigma')^{3}}-3\dfrac{V''_{E}\sigma''}{V_{E}(\sigma')^{4}}+3\dfrac{V'_{E}(\sigma'')^{2}}{V_{E}(\sigma')^{5}}-\dfrac{V'_{E}\sigma'''}{V_{E}(\sigma')^{4}}\right),
\end{eqnarray}
where the prime denotes a derivative with respect to $\phi$, while
$\lambda(\phi)$ is taken as $\lambda(\phi_{0})$. The scalar spectral
index, tensor-to-scalar ratio, running of the scalar spectral index,
and power spectrum~\cite{Lyth:1998xn-1} are respectively
\begin{eqnarray}
n_{s} & = & 1-6\epsilon+2\eta+2\left[\frac{1}{3}\eta^{2}-\left(\frac{5}{3}+12C\right)\epsilon^{2}+(8C-1)\epsilon\eta-(C-\frac{1}{3})\zeta\right]~,\\
r & = & 16\epsilon\left[1+\frac{2}{3}(3C-1)(2\epsilon-\eta)\right]~,\\
\alpha_{s} & = & \frac{\dd n_{s}}{\dd\ln k}=16\epsilon\eta-24\epsilon^{2}-2\zeta~,~\\
P_{s} & = & \dfrac{V_{E}}{24\pi^{2}\epsilon},
\end{eqnarray}
where $C=-2+\ln2+\gamma\simeq-0.7296$ with $\gamma$ the Euler\textendash Mascheroni
constant. Here, we also considered the second-order corrections, which
will give extra contributions of $\delta n_{s}\sim0.001$ and $\delta r\sim-1\times10^{-4}$.
The e-folding number is given by
\begin{eqnarray}
N & = & \frac{1}{M_{{\rm Pl}}}\int_{\phi_{e}}^{\phi_{0}}\dfrac{\dd\phi}{\sqrt{2\epsilon(\phi)}}\left(\dfrac{\dd\sigma}{\dd\phi}\right)
\end{eqnarray}
and the inflation ends once $\epsilon(\phi_{e})$ or $|\eta(\phi_{e})|$ reaches 1.

In the following numerical discussions, we shall find that inflation
always ends when $\epsilon=1$ in Models I and II. So let us study
the corresponding constraint on $\xi$. For $\epsilon(\phi_{e})=1$,
there are two positive roots and two negative roots for $\phi_{e}$: 
\begin{eqnarray}
 &  & \phi_{1}=\frac{4}{\sqrt{1+\sqrt{(8\xi+1)(24\xi+1)}}}~,~\label{df}\\
 &  & \phi_{2}=\frac{4}{\sqrt{1-\sqrt{(8\xi+1)(24\xi+1)}}}~,~\\
 &  & \phi_{3}=-\frac{4}{\sqrt{1+\sqrt{(8\xi+1)(24\xi+1)}}}~,~
 \end{eqnarray}
 \begin{eqnarray}
 &  & \phi_{4}=-\frac{4}{\sqrt{1-\sqrt{(8\xi+1)(24\xi+1)}}}.
\end{eqnarray}
Because we require both $\phi>0$ and $\xi>0$, Eq.~(\ref{df})
is the only physical solution. Let us study the possible value of
$\xi$ in the following

(1) Assume that $\xi$ locates in the range of $0<\xi\leq1$, we
obtain $\phi_{e}\geq2\sqrt{\frac{2}{7}}~M_{{\rm Pl}}$ from Eq.~(\ref{df}).
And then inflation ends above the reduced Planck scale, Thus, we can
have $0<\xi\leq1$ only for the trans-Planckian inflation.

\begin{figure}[H]
\centering
\includegraphics[height=4cm]{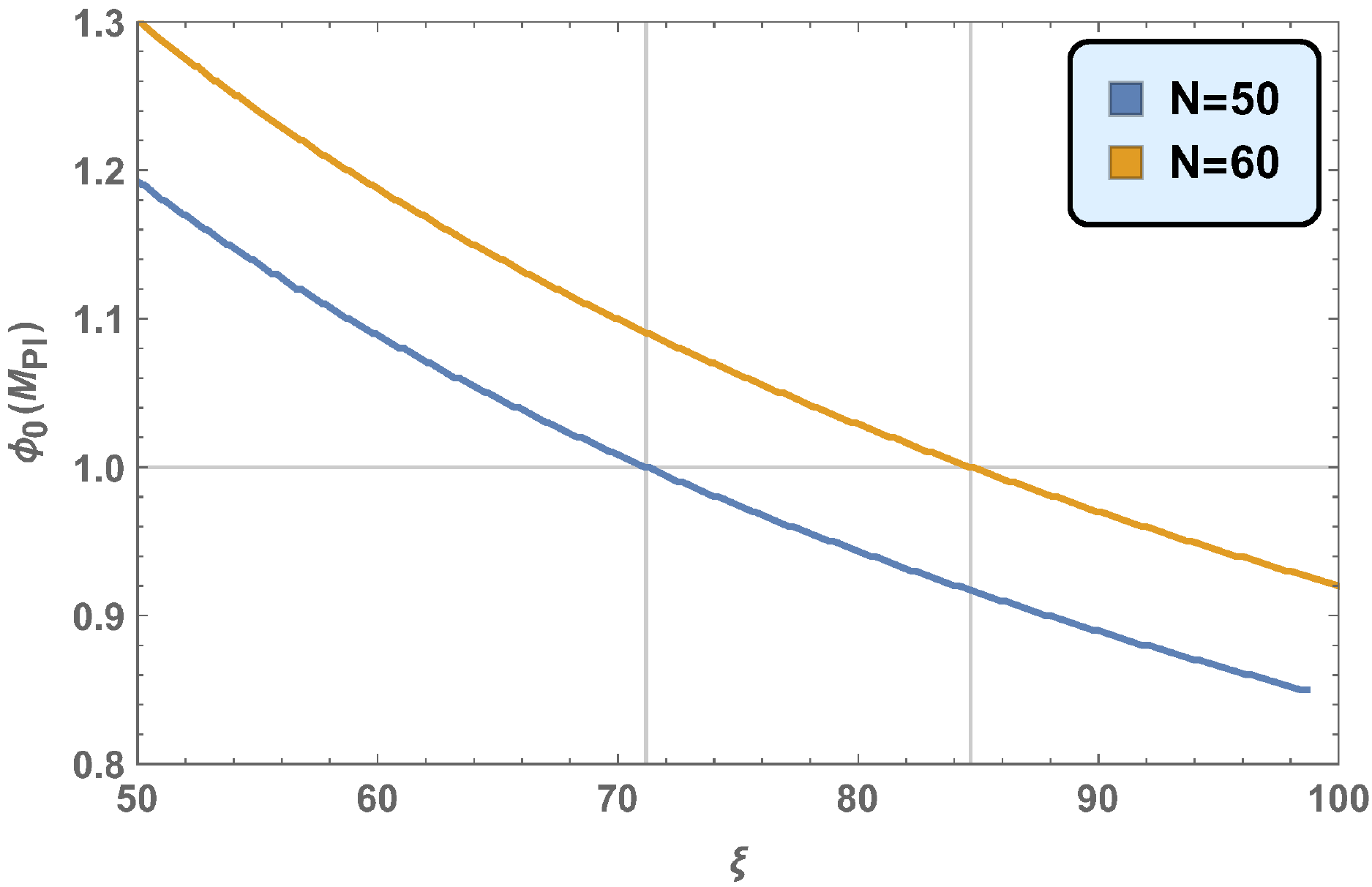}
\caption{The gray lines is corresponding to $\xi=71.16$ with $N=50$ and $\xi=84.67$
with $N=60$.}
\label{fig:xi_phi0} 
\end{figure}

(2) To satisfy the condition $0<\phi_{e}<M_{{\rm Pl}}$, we find
that $\xi$ can only locate in the ranges of $\xi>1$ or $\xi<-\frac{7}{6}$.
With $\xi>1$ and $\phi_{0}>\phi_{e}$, the number of e-folding is
\begin{equation}
N=\frac{1}{8}\left(6\log\left(\frac{1+\xi\phi_{e}^{2}}{1+\xi\phi_{0}^{2}}\right)+(1+6\xi)(\phi_{0}-\phi_{e})(\phi_{0}+\phi_{e})\right)\label{n}
\end{equation}
With $N$ fixed, Eq.~(\ref{n}) is actually a functional relation
between $\xi$ and $\phi_{0}$. We present $\xi$ versus $\phi_{0}$
with the fixed $N$ in Fig.~\ref{fig:xi_phi0}. To have $\phi_{0}\leq M_{{\rm Pl}}$,
we obtain $\xi\geq71.16$ for $N=50$ and $\xi\geq84.67$ for $N=60$
from Fig.~\ref{fig:xi_phi0}.

In short, for Higgs inflation taking place below the reduced Planck
scale, \textit{{}i.e.}{}, the sub-Planckian inflation, there
is a lower bound on $\xi$, which is $71.16$ for $N=50$ and $84.67$
for $N=60$. Especially, it is impossible to get $0<\xi\leq1$. This
conclusion is model independent as long as the Higgs-gravity coupling
is taken to be in Eq.~(\ref{S}).

\section{Numerical Study for Inflation}

We will discuss the numerical results for the inflationary observables
as our predictions, which can fit the current experimental data. We
define the precise gauge coupling unification condition as $\alpha_{1}^{-1}\equiv(\alpha_{2}^{-1}+\alpha_{3}^{-1})/2$,
as well as consider the string scale unification with $M_{U}=3\times10^{17}{\rm ~GeV}$
and the reduced Planck scale unification with $M_{U}=M_{{\rm Pl}}=2.43\times10^{18}{\rm ~GeV}$.

{For a fixed $N=50$ and 60, we find that the inflationary predictions
or observables can be consistent with all the current experimental
constraints from the Planck, Baryon Acoustic Oscillations (BAO), and
BICEP2/Keck Array data, etc~\cite{Ade:2015lrj,Array:2015xqh}. And
the couplings $\xi$ can be adjusted to realize the observed power
spectrum $P_{s}=2.20\times10^{-9}$. In particular, we find that inflation
always ends when $\epsilon=1$ in our models. For numerical results,
we present some concrete benchmark points for Model I in Tables~\ref{tab:small_xi_case1}
as well as for Model II in Tables~\ref{tab:small_xi_case2}. To achieve
the gauge coupling unification at the reduced Planck scale, we find
that $M_{8}\sim1.0\times10^{6}~{\rm GeV}$ and $M_{3}\sim1.0\times10^{12}~{\rm GeV}$
in Model I, as well as $M_{8}\sim1.25\times10^{9}~{\rm GeV}$ and
$M_{3}\sim1.24\times10^{9}~{\rm GeV}$ in Model II. Thus, we have
$M_{8}\simeq M_{3}$ in Model II, which makes it much more interesting.
Neglecting $G_{x}$ and $W_{x}$ in Model II, we do have gauge coupling
unification around $2\times10^{16}$~GeV~\cite{Chen:2017rpn}. Thus,
introducing the $SU(3)_{C}$ and $SU(2)_{L}$ adjoint fermions/chiral
superfields with the similar intermediate-scale mass in the GUTs,
supersymmetric GUTs and string models, we can lift the tradition GUT
scale to the string scale or reduced Planck scale in general, which
will be very important at least in the string model building! Moreover,
the scalar spectral indices are almost the same and they are around
0.9626 and 0.9685 respectively for $N=50$ and $N=60$, the tensor-to-scalar
ratios are at the order of $10^{-3}$, and the running of the scalar
spectral index is negative and at the order of $10^{-6}$. In particular,
with some fine-tuning of the input parameters, we can obtain that
$\lambda(\phi_{0})$ is at the order of $10^{-6}$, and then $\xi$s
are close to their lower bounds around $\xi\sim71-84$, which are
given in Tables~\ref{tab:small_xi_case1} and \ref{tab:small_xi_case2}.
In short, we indeed solve the $\xi$ problem in the Higgs inflation.
In Tables }\ref{tab:small_xi_case1_a} - \ref{tab:small_xi_case2_b},
the values at the center of both tables is the Seesaw scale $M_{R}$
adopted to meet requirements. $\lambda$ should be as small as possible
for the purpose of Higgs inflation happening at inflation scale but
still above zero to solve the stability problem. The precision of
$M_{R}$ shown in the Tables \ref{tab:small_xi_case1_a}{ - \ref{tab:small_xi_case2_b}
is due to the small $\zeta$ in our models. ~\ref{tab:small_xi_case1_a}
and \ref{tab:small_xi_case2_a} are the allowed range of paramters
of $m_{t}$ and $M_{h}$ in model I when the unification is around
the reduced Planck scale. Tables \ref{tab:small_xi_case1_b}and \ref{tab:small_xi_case2_b}
are for model II when the unification happens at the string scale.}

\begin{figure}[H]
	\centering \subfigure[]{\includegraphics[height=3.2cm]{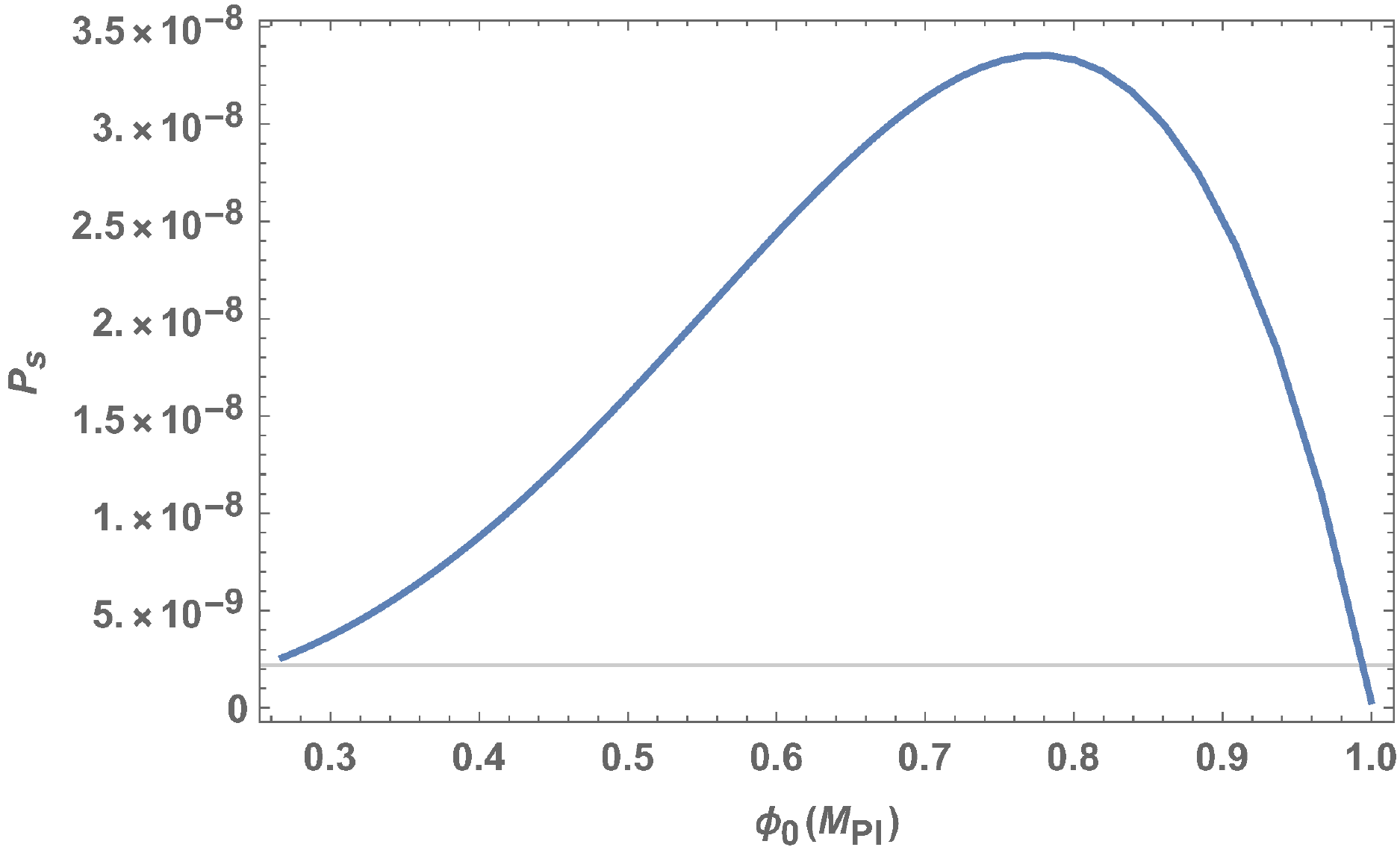}}
	\subfigure[]{\includegraphics[height=3.2cm]{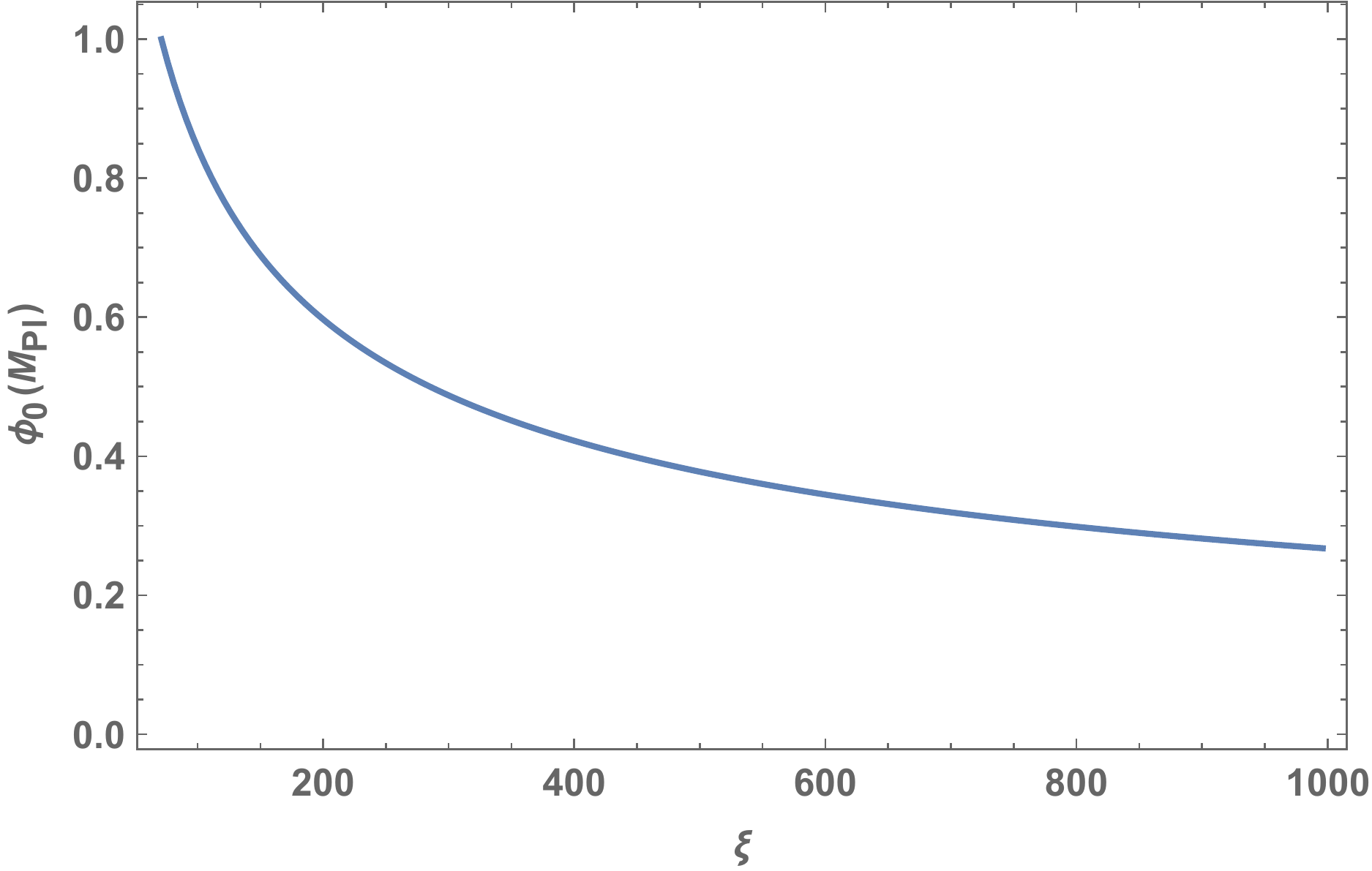}} \subfigure[]{\includegraphics[height=3.2cm]{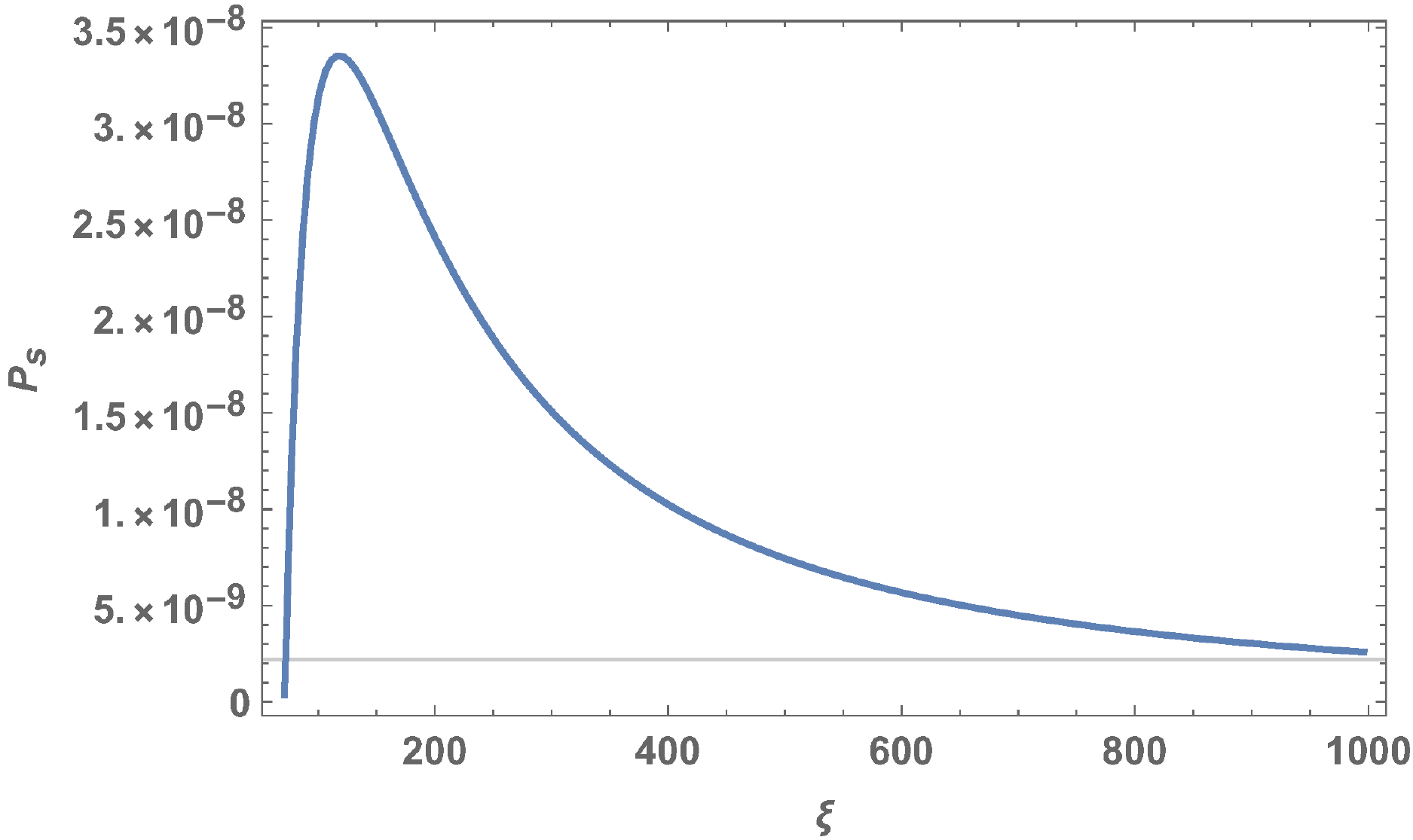}}
	\caption{ The relation between $P_{s}$, $\phi_{0}$ and $\xi$ for Model II
		with $m_{t}=173.34{\rm ~GeV}$,$M_{h}=125.09{\rm ~GeV}$ and $M_U=2.43\times10^{18}{\rm ~GeV}$. The e-folding number is fixed at 50.}
	\label{fig:ps_62} 
\end{figure}

To understand our numerical results with small $\xi$ close to its
lower bound, let us study it in more details. The scalar power spectrum
amplitude is given by
\begin{eqnarray}
P_{s}(\xi,\phi_{0}) & = & \dfrac{1}{24\pi^{2}}\dfrac{V(\xi,\phi_{0})}{\epsilon(\xi,\phi_{0})}\nonumber \\
 & = & \dfrac{\lambda(\phi_{0})\phi_{0}^{6}}{1536\pi^{2}}\times\dfrac{1+\xi(1+6\xi)\phi_{0}^{2}}{(1+\xi\phi_{0}^{2})^{2}},
\end{eqnarray}
where the Higgs coupling $\lambda(\phi_{0})$ also depends on the
parameters involved in the RGE evolution, $\lambda(\phi_{0})=\lambda(\phi_{0}|m_{t},M_{R},M_{8},M_{3},M_{U})$.
With $\xi$ given and $N=50$ fixed, $\phi_{0}$ is determined, so
we actually have $\phi_{0}=\phi_{0}(\xi)$. The relations between
$(\xi,\phi_{0},P_{s})$ are plotted as the functions $\phi_{0}=\phi_{0}(\xi)$,
$P_{s}=P_{s}(\phi_{0}|m_{t},M_{R},M_{8},M_{3},M_{U})$, and $P_{s}=P_{s}(\xi|m_{t},M_{R},M_{8},M_{3},M_{U})$
for Model II in Figs.~\ref{fig:ps_62}, where $(m_{t},M_{R},M_{8},M_{3},M_{U})$
are taken to be $(173.34,1.7786\times10^{14},1.25\times10^{9},1.24\times10^{9},2.43\times10^{18})$~GeV.
As we can see from these figures, $\phi_{0}=\phi_{0}(\xi)$ is an
input parameter independent function, however, $\phi_{0}$ decreases when $\xi$
increases. On the other hand, $P_{s}=P_{s}(\phi_{0}|m_{t},M_{R},M_{8},M_{3},M_{U})$
and $P_{s}=P_{s}(\xi|m_{t},M_{R},M_{8},M_{3},M_{U})$ delicately depend
on the selected input parameters. With little changes in $m_{t}$
and $M_{R}$, the functions with the local maximum become the functions
with the local minimum. The gray lines in Fig.~\ref{fig:ps_62} give the constraint of the correct power spectrum $P_{s}=2.20\times10^{-9}$.
Thus, we show explicitly that there are two viable regions in the
$P_{s}$ versus $\phi_{0}(M_{{\rm Pl}})$ and $P_{s}$ versus $\xi$
planes, which give the observed power spectrum. One kind of viable
regions has small $\xi\subset(70-86)$ and large $\phi_{0}\sim M_{{\rm Pl}}$,
and the benchmark points are given in Tables \ref{tab:small_xi_case1} and
\ref{tab:small_xi_case2}. In both of our models, we can indeed obtain
$\xi$ around $70-80$, which is almost the minimum $\xi$ that can
be realized. And then the $\xi$ problem is indeed solved.
 
$\;\\$
 
$\;$
 
$\;$
 
$\;$
 
$\;\\$
 
$\;$ 
\begin{table}[H]
	{\raggedright{}}%
	\begin{tabular}{|c|c|c|c|c|c|c|c|c|}
		\hline
		& \multicolumn{8}{c|}{{\footnotesize{}{}{}$m_{t}=173.34$,$M_{h}=125.09$,$M_{R}=1.7786\times10^{14}$,
				$M_{8}=1.\times10^{6}$, $M_{3}=1.13\times10^{12}$, $M_{U}=2.43\times10^{18}$}}\tabularnewline
		\hline
		{\footnotesize{}{}{}N}  & {\footnotesize{}{}{}$\xi$ }  & {\footnotesize{}{}{}$\phi_{0}({\rm M_{Pl})}$}  & {\footnotesize{}{}{}$\phi_{e}({\rm M_{Pl})}$}  & {\footnotesize{}{}{}$n_{s}$}  & {\footnotesize{}{}{}$r(10^{-3})$}  & {\footnotesize{}{}{}$\alpha(10^{-4})$}  & {\footnotesize{}{}{}$\lambda(\phi_{0})$}  & {\footnotesize{}{}{}$\lambda(M_{U})$}\tabularnewline
		\hline
		{\footnotesize{}{}{}50}  & {\footnotesize{}{}{}145.8}  & {\footnotesize{}{}{}0.6991}  & {\footnotesize{}{}{}0.0889}  & {\footnotesize{}{}{}0.9626}  & {\footnotesize{}{}{}4.03}  & {\footnotesize{}{}{}-7.48}  & {\footnotesize{}{}{}$2.39\times10^{-5}$}  & {\footnotesize{}{}{}$5.36\times10^{-5}$}\tabularnewline
		\hline
		{\footnotesize{}{}{}60}  & {\footnotesize{}{}{}173.7}  & {\footnotesize{}{}{}0.6985}  & {\footnotesize{}{}{}0.0815}  & {\footnotesize{}{}{}0.9685}  & {\footnotesize{}{}{}2.87}  & {\footnotesize{}{}{}-5.23}  & {\footnotesize{}{}{}$2.39\times10^{-5}$}  & {\footnotesize{}{}{}$5.36\times10^{-5}$ }\tabularnewline
		\hline
		\hline
		& \multicolumn{8}{c|}{{\footnotesize{}{}{}$m_{t}=172.58$,$M_{h}=126.89$,$M_{R}=2.1286\times10^{14}$,
				$M_{8}=2.3\times10^{9}$, $M_{3}=2.1\times10^{14}$, $M_{U}=3.0\times10^{17}$}}\tabularnewline
		\hline
		{\footnotesize{}{}{}N}  & {\footnotesize{}{}{}$\xi$ }  & {\footnotesize{}{}{}$\phi_{0}({\rm M_{Pl})}$}  & {\footnotesize{}{}{}$\phi_{e}({\rm M_{Pl})}$}  & {\footnotesize{}{}{}$n_{s}$}  & {\footnotesize{}{}{}$r(10^{-3})$}  & {\footnotesize{}{}{}$\alpha(10^{-4})$}  & {\footnotesize{}{}{}$\lambda(\phi_{0})$}  & {\footnotesize{}{}{}$\lambda(M_{U})$}\tabularnewline
		\hline
		{\footnotesize{}{}{}50}  & {\footnotesize{}{}{}73.05}  & {\footnotesize{}{}{}0.98706}  & {\footnotesize{}{}{}0.1256}  & {\footnotesize{}{}{}0.9626}  & {\footnotesize{}{}{}4.03}  & {\footnotesize{}{}{}-7.48}  & {\footnotesize{}{}{}$6.01\times10^{-6}$}  & {\footnotesize{}{}{}$3.29\times10^{-3}$}\tabularnewline
		\hline
		{\footnotesize{}{}{}60}  & {\footnotesize{}{}{}86.91}  & {\footnotesize{}{}{}0.98707}  & {\footnotesize{}{}{}0.1152}  & {\footnotesize{}{}{}0.9685}  & {\footnotesize{}{}{}2.87}  & {\footnotesize{}{}{}-5.23}  & {\footnotesize{}{}{}$6.00\times10^{-6}$}  & {\footnotesize{}{}{}$3.29\times10^{-3}$ }\tabularnewline
		\hline
		& \multicolumn{8}{c|}{{\footnotesize{}{}{}$m_{t}=172.58$,$M_{h}=126.89$, $M_{R}=2.4644\times10^{14}$,
				$M_{8}=1.07\times10^{6}$, $M_{3}=1.\times10^{12}$, $M_{U}=2.43\times10^{18}$}}\tabularnewline
		\hline
		{\footnotesize{}{}{}N}  & {\footnotesize{}{}{}$\xi$ }  & {\footnotesize{}{}{}$\phi_{0}({\rm M_{Pl})}$}  & {\footnotesize{}{}{}$\phi_{e}({\rm M_{Pl})}$}  & {\footnotesize{}{}{}$n_{s}$}  & {\footnotesize{}{}{}$r(10^{-3})$}  & {\footnotesize{}{}{}$\alpha(10^{-4})$}  & {\footnotesize{}{}{}$\lambda(\phi_{0})$}  & {\footnotesize{}{}{}$\lambda(M_{U})$}\tabularnewline
		\hline
		{\footnotesize{}{}{}50}  & {\footnotesize{}{}{}71.237}  & {\footnotesize{}{}{}0.9995}  & {\footnotesize{}{}{}0.1272}  & {\footnotesize{}{}{}0.9626}  & {\footnotesize{}{}{}4.03}  & {\footnotesize{}{}{}-7.48 }  & {\footnotesize{}{}{}$5.72\times10^{-6}$}  & {\footnotesize{}{}{}$2.11\times10^{-6}$}\tabularnewline
		\hline
		{\footnotesize{}{}{}60}  & {\footnotesize{}{}{}84.763}  & {\footnotesize{}{}{}0.9995}  & {\footnotesize{}{}{}0.1166}  & {\footnotesize{}{}{}0.9685}  & {\footnotesize{}{}{}2.87}  & {\footnotesize{}{}{}-5.23}  & {\footnotesize{}{}{}$5.69\times10^{-6}$}  & {\footnotesize{}{}{}$2.11\times10^{-6}$ }\tabularnewline
		\hline
		\hline
		& \multicolumn{8}{c|}{{\footnotesize{}{}{}$m_{t}=174.10$,$M_{h}=126.89$, $M_{R}=1.8374\times10^{14}$,
				$M_{8}=1.\times10^{6}$, $M_{3}=1.12\times10^{12}$, $M_{U}=2.43\times10^{18}$}}\tabularnewline
		\hline
		{\footnotesize{}{}{}N}  & {\footnotesize{}{}{}$\xi$ }  & {\footnotesize{}{}{}$\phi_{0}({\rm M_{Pl})}$}  & {\footnotesize{}{}{}$\phi_{e}({\rm M_{Pl})}$}  & {\footnotesize{}{}{}$n_{s}$}  & {\footnotesize{}{}{}$r(10^{-3})$}  & {\footnotesize{}{}{}$\alpha(10^{-4})$}  & {\footnotesize{}{}{}$\lambda(\phi_{0})$}  & {\footnotesize{}{}{}$\lambda(M_{U})$}\tabularnewline
		\hline
		{\footnotesize{}{}{}50}  & {\footnotesize{}{}{}72.78 }  & {\footnotesize{}{}{}0.9889}  & {\footnotesize{}{}{}0.1258}  & {\footnotesize{}{}{}0.9626}  & {\footnotesize{}{}{}4.03}  & {\footnotesize{}{}{}-7.48}  & {\footnotesize{}{}{}$5.97\times10^{-6}$}  & {\footnotesize{}{}{}$4.72\times10^{-6}$}\tabularnewline
		\hline
		{\footnotesize{}{}{}60}  & {\footnotesize{}{}{}86.53}  & {\footnotesize{}{}{}0.9892}  & {\footnotesize{}{}{}0.1154}  & {\footnotesize{}{}{}0.9685}  & {\footnotesize{}{}{}2.87}  & {\footnotesize{}{}{}-5.23}  & {\footnotesize{}{}{}$5.93\times10^{-6}$}  & {\footnotesize{}{}{}$4.72\times10^{-6}$ }\tabularnewline
		\hline
	\end{tabular}
	\caption{The benchmark points with small $\xi$ in Model I for $N=50$ and
		60, where the mass unit is  GeV.}
	\label{tab:small_xi_case1} 
\end{table}

\begin{table}[H]
{\raggedright{}}%
\begin{tabular}{|c|c|c|c|c|c|c|c|c|}
\hline
 & \multicolumn{8}{c|}{{\footnotesize{}{}{}$m_{t}=173.34$,$M_{h}=125.09$,$M_{R}=1.9963\times10^{14}$,
$M_{8}=1.02\times10^{12}$,$M_{3}=1.02\times10^{12}$,$M_{U}=3.0\times10^{17}$}}\tabularnewline
\hline
{\footnotesize{}{}{}N}  & {\footnotesize{}{}{}$\xi$ }  & {\footnotesize{}{}{}$\phi_{0}({\rm M_{Pl})}$}  & {\footnotesize{}{}{}$\phi_{e}({\rm M_{Pl})}$}  & {\footnotesize{}{}{}$n_{s}$}  & {\footnotesize{}{}{}$r(10^{-3})$}  & {\footnotesize{}{}{}$\alpha(10^{-4})$}  & {\footnotesize{}{}{}$\lambda(\phi_{0})$}  & {\footnotesize{}{}{}$\lambda(M_{U})$}\tabularnewline
\hline
{\footnotesize{}{}{}50}  & {\footnotesize{}{}{}74.442}  & {\footnotesize{}{}{}0.9778}  & {\footnotesize{}{}{}0.1244}  & {\footnotesize{}{}{}0.9626}  & {\footnotesize{}{}{}4.03}  & {\footnotesize{}{}{}-7.48}  & {\footnotesize{}{}{}$6.24\times10^{-6}$}  & {\footnotesize{}{}{}$1.41\times10^{-3}$}\tabularnewline
\hline
{\footnotesize{}{}{}60}  & {\footnotesize{}{}{}88.558}  & {\footnotesize{}{}{}0.9779}  & {\footnotesize{}{}{}0.1141}  & {\footnotesize{}{}{}0.9685}  & {\footnotesize{}{}{}2.87}  & {\footnotesize{}{}{}-5.23}  & {\footnotesize{}{}{}$6.21\times10^{-6}$}  & {\footnotesize{}{}{}$1.41\times10^{-3}$}\tabularnewline
\hline
 & \multicolumn{8}{c|}{{\footnotesize{}{}{}$m_{t}=173.34$,$M_{h}=125.09$,$M_{R}=2.2135\times10^{14}$,$M_{8}=1.24\times10^{9}$,$M_{3}=1.24\times10^{9}$,$M_{U}=2.43\times10^{18}$}}\tabularnewline
\hline
{\footnotesize{}{}{}N}  & {\footnotesize{}{}{}$\xi$ }  & {\footnotesize{}{}{}$\phi_{0}({\rm M_{Pl})}$}  & {\footnotesize{}{}{}$\phi_{e}({\rm M_{Pl})}$}  & {\footnotesize{}{}{}$n_{s}$}  & {\footnotesize{}{}{}$r(10^{-3})$}  & {\footnotesize{}{}{}$\alpha(10^{-4})$}  & {\footnotesize{}{}{}$\lambda(\phi_{0})$}  & {\footnotesize{}{}{}$\lambda(M_{U})$}\tabularnewline
\hline
{\footnotesize{}{}{}50}  & {\footnotesize{}{}{}71.923}  & {\footnotesize{}{}{}0.9947}  & {\footnotesize{}{}{}0.1266}  & {\footnotesize{}{}{}0.9626}  & {\footnotesize{}{}{}4.03}  & {\footnotesize{}{}{}-7.48}  & {\footnotesize{}{}{}$5.83\times10^{-6}$}  & {\footnotesize{}{}{}$1.14\times10^{-6}$}\tabularnewline
\hline
{\footnotesize{}{}{}60}  & {\footnotesize{}{}{}85.566}  & {\footnotesize{}{}{}0.9947}  & {\footnotesize{}{}{}0.1160}  & {\footnotesize{}{}{}0.9685}  & {\footnotesize{}{}{}2.87}  & {\footnotesize{}{}{}-5.23}  & {\footnotesize{}{}{}$5.80\times10^{-5}$}  & {\footnotesize{}{}{}$1.14\times10^{-6}$ }\tabularnewline
\hline
\hline
 & \multicolumn{8}{c|}{{\footnotesize{}{}{}$m_{t}=172.58$,$M_{h}=126.89$,$M_{R}=2.6295\times10^{14}$,$M_{8}=1.02\times10^{12}$,$M_{3}=1.02\times10^{12}$,$M_{U}=3.0\times10^{17}$}}\tabularnewline
\hline
{\footnotesize{}{}{}N}  & {\footnotesize{}{}{}$\xi$ }  & {\footnotesize{}{}{}$\phi_{0}({\rm M_{Pl})}$}  & {\footnotesize{}{}{}$\phi_{e}({\rm M_{Pl})}$}  & {\footnotesize{}{}{}$n_{s}$}  & {\footnotesize{}{}{}$r(10^{-3})$}  & {\footnotesize{}{}{}$\alpha(10^{-4})$}  & {\footnotesize{}{}{}$\lambda(\phi_{0})$}  & {\footnotesize{}{}{}$\lambda(M_{U})$}\tabularnewline
\hline
{\footnotesize{}{}{}50}  & {\footnotesize{}{}{}76.157}  & {\footnotesize{}{}{}0.9668}  & {\footnotesize{}{}{}0.1230}  & {\footnotesize{}{}{}0.9625}  & {\footnotesize{}{}{}4.03}  & {\footnotesize{}{}{}-7.48 }  & {\footnotesize{}{}{}$6.53\times10^{-5}$}  & {\footnotesize{}{}{}$6.58\times10^{-3}$}\tabularnewline
\hline
{\footnotesize{}{}{}60}  & {\footnotesize{}{}{}87.223}  & {\footnotesize{}{}{}0.9853}  & {\footnotesize{}{}{}0.1150}  & {\footnotesize{}{}{}0.9685}  & {\footnotesize{}{}{}2.87}  & {\footnotesize{}{}{}-5.23}  & {\footnotesize{}{}{}$6.02\times10^{-6}$}  & {\footnotesize{}{}{}$6.58\times10^{-3}$ }\tabularnewline
\hline
 & \multicolumn{8}{c|}{{\footnotesize{}{}{}$m_{t}=172.58$,$M_{h}=126.89$,$M_{R}=2.7911\times10^{14}$,$M_{8}=1.24\times10^{9}$,$M_{3}=1.24\times10^{9}$,$M_{U}=2.43\times10^{18}$}}\tabularnewline
\hline
{\footnotesize{}{}{}N}  & {\footnotesize{}{}{}$\xi$ }  & {\footnotesize{}{}{}$\phi_{0}({\rm M_{Pl})}$}  & {\footnotesize{}{}{}$\phi_{e}({\rm M_{Pl})}$}  & {\footnotesize{}{}{}$n_{s}$}  & {\footnotesize{}{}{}$r(10^{-3})$}  & {\footnotesize{}{}{}$\alpha(10^{-4})$}  & {\footnotesize{}{}{}$\lambda(\phi_{0})$}  & {\footnotesize{}{}{}$\lambda(M_{U})$}\tabularnewline
\hline
{\footnotesize{}{}{}50}  & {\footnotesize{}{}{}71.383}  & {\footnotesize{}{}{}0.9985}  & {\footnotesize{}{}{}0.1270}  & {\footnotesize{}{}{}0.9625}  & {\footnotesize{}{}{}4.03}  & {\footnotesize{}{}{}-7.48 }  & {\footnotesize{}{}{}$5.74\times10^{-6}$}  & {\footnotesize{}{}{}$1.05\times10^{-6}$}\tabularnewline
\hline
{\footnotesize{}{}{}60}  & {\footnotesize{}{}{}84.927}  & {\footnotesize{}{}{}0.9985}  & {\footnotesize{}{}{}0.1165}  & {\footnotesize{}{}{}0.9685}  & {\footnotesize{}{}{}2.87}  & {\footnotesize{}{}{}-5.23}  & {\footnotesize{}{}{}$5.71\times10^{-6}$}  & {\footnotesize{}{}{}$1.05\times10^{-6}$ }\tabularnewline
\hline
\hline
 & \multicolumn{8}{c|}{{\footnotesize{}{}{}$m_{t}=174.10$,$M_{h}=126.89$,$M_{R}=2.0542\times10^{14}$,$M_{8}=1.02\times10^{12}$,$M_{3}=1.02\times10^{12}$,$M_{U}=3.0\times10^{17}$}}\tabularnewline
\hline
{\footnotesize{}{}{}N}  & {\footnotesize{}{}{}$\xi$ }  & {\footnotesize{}{}{}$\phi_{0}({\rm M_{Pl})}$}  & {\footnotesize{}{}{}$\phi_{e}({\rm M_{Pl})}$}  & {\footnotesize{}{}{}$n_{s}$}  & {\footnotesize{}{}{}$r(10^{-3})$}  & {\footnotesize{}{}{}$\alpha(10^{-4})$}  & {\footnotesize{}{}{}$\lambda(\phi_{0})$}  & {\footnotesize{}{}{}$\lambda(M_{U})$}\tabularnewline
\hline
{\footnotesize{}{}{}50}  & {\footnotesize{}{}{}73.773}  & {\footnotesize{}{}{}0.9822}  & {\footnotesize{}{}{}0.1250}  & {\footnotesize{}{}{}0.9626}  & {\footnotesize{}{}{}4.03}  & {\footnotesize{}{}{}-7.48}  & {\footnotesize{}{}{}$6.13\times10^{-6}$}  & {\footnotesize{}{}{}$1.89\times10^{-3}$}\tabularnewline
\hline
{\footnotesize{}{}{}60}  & {\footnotesize{}{}{}87.765}  & {\footnotesize{}{}{}0.9823}  & {\footnotesize{}{}{}0.1146}  & {\footnotesize{}{}{}0.9685}  & {\footnotesize{}{}{}2.87}  & {\footnotesize{}{}{}-5.23}  & {\footnotesize{}{}{}$6.10\times10^{-6}$}  & {\footnotesize{}{}{}$1.89\times10^{-3}$ }\tabularnewline
\hline
 & \multicolumn{8}{c|}{{\footnotesize{}{}{}$m_{t}=174.10$,$M_{h}=126.89$,$M_{R}=2.2679\times10^{14}$,$M_{8}=1.24\times10^{9}$,$M_{3}=1.24\times10^{9}$,$M_{U}=2.43\times10^{18}$}}\tabularnewline
\hline
{\footnotesize{}{}{}N}  & {\footnotesize{}{}{}$\xi$ }  & {\footnotesize{}{}{}$\phi_{0}({\rm M_{Pl})}$}  & {\footnotesize{}{}{}$\phi_{e}({\rm M_{Pl})}$}  & {\footnotesize{}{}{}$n_{s}$}  & {\footnotesize{}{}{}$r(10^{-3})$}  & {\footnotesize{}{}{}$\alpha(10^{-4})$}  & {\footnotesize{}{}{}$\lambda(\phi_{0})$}  & {\footnotesize{}{}{}$\lambda(M_{U})$}\tabularnewline
\hline
{\footnotesize{}{}{}50}  & {\footnotesize{}{}{}71.875}  & {\footnotesize{}{}{}0.9951}  & {\footnotesize{}{}{}0.1266}  & {\footnotesize{}{}{}0.9626}  & {\footnotesize{}{}{}4.03}  & {\footnotesize{}{}{}-7.48}  & {\footnotesize{}{}{}$5.81\times10^{-6}$}  & {\footnotesize{}{}{}$3.95\times10^{-7}$}\tabularnewline
\hline
{\footnotesize{}{}{}60}  & {\footnotesize{}{}{}85.510}  & {\footnotesize{}{}{}0.9951}  & {\footnotesize{}{}{}0.1161}  & {\footnotesize{}{}{}0.9685}  & {\footnotesize{}{}{}2.87}  & {\footnotesize{}{}{}-5.23}  & {\footnotesize{}{}{}$5.79\times10^{-6}$}  & {\footnotesize{}{}{}$3.95\times10^{-7}$ }\tabularnewline
\hline
\end{tabular}
\caption{The benchmark points with small $\xi$ in Model II for $N=50$ and
60, where the masses are  in GeV.\label{tab:small_xi_case2} }
\end{table}

\newpage
\begin{table}[H]
\centering %
\begin{tabular}{|c|ccccc|c|}
\hline
\diagbox{$m_{t}$}{$M_{h}$}  & 121.49  & 123.29  & 125.09  & 126.89  & 128.69  & Mini\tabularnewline
\hline
171.82  & -  & $~1.6510\times10^{14}~$  & $~2.0658\times10^{14}~$  & $~2.4343\times10^{14}~$  & $~2.7766\times10^{14}~$  & 122.70\tabularnewline
172.58  & -  & -  & $~1.7088\times10^{14}~$  & $~2.1288\times10^{14}~$  & $~2.5041\times10^{14}~$  & 124.17\tabularnewline
173.34  & -  & -  & -  & $~1.7688\times10^{14}~$  & $~2.1950\times10^{14}~$  & 125.65\tabularnewline
174.10  & -  & -  & -  & -  & $~2.2390\times10^{14}~$  & 127.12\tabularnewline
174.86  & -  & -  & -  & -  & $~1.1929\times10^{14}~$  & 128.60\tabularnewline
\hline
\end{tabular}\caption{For \textbf{Model I}, dependence of right handed neutrino mass on the top and SM Higgs masses when gauge couplings are unified at  $M_{U}=3.0\times10^{17}~{\rm GeV}$.
We use hyphen when there is no solution for given Higgs and top quark satisfying vacuum stability condition. For all solutions presented here  the Higgs quartic coupling is $O(10^{-6})$ at $M_U$ scale. The last column represents  the lowest bounds of $M_{h}$ which satisfies vacuum stability bound for give top quark mass.}
\label{tab:small_xi_case1_b}
\end{table}

\begin{table}[H]
\centering %
\begin{tabular}{|c|ccccc|c|}
\hline
\diagbox{$m_{t}$}{$M_{h}$}  & 121.49  & 123.29  & 125.09  & 126.89  & 128.69  & Mini\tabularnewline
\hline
171.82  & $~1.8920\times10^{14}~$  & $~2.2502\times10^{14}~$  & $~2.5649\times10^{14}~$  & $~2.8919\times10^{14}~$  & $~3.1930\times10^{14}~$  & 121.49\tabularnewline
172.58  & -  & $~1.9418\times10^{14}~$  & $~2.2992\times10^{14}~$  & $~2.6295\times10^{14}~$  & $~2.9430\times10^{14}~$  & 122.42\tabularnewline
173.34  & -  & -  & $~1.9963\times10^{14}~$  & $~2.3604\times10^{14}~$  & $~2.6976\times10^{14}~$  & 123.86\tabularnewline
174.10  & -  & -  & -  & $~2.0542\times10^{14}~$  & $~2.4251\times10^{14}~$  & 125.30\tabularnewline
174.86  & -  & -  & -  & $~1.6689\times10^{14}~$  & $~2.1157\times10^{14}~$  & 126.75\tabularnewline
\hline
\end{tabular}\caption{For \textbf{Model II}, dependence of right handed neutrino mass on the top and SM Higgs masses when gauge couplings are unified at  $M_{U}=3.0\times10^{17}~{\rm GeV}$.
We use hyphen when there is no solution for given Higgs and top quark satisfying vacuum stability condition. For all solutions presented here  the Higgs quartic coupling is $O(10^{-6})$ at $M_U$ scale. The last column represents  the lowest bounds of $M_{h}$ which satisfies vacuum stability bound for give top quark mass.}
\label{tab:small_xi_case2_b}
\end{table}


\section{Conclusion}

We have studied two non-supersymmetric models with gauge couplings unification. 
In these scenarios, the so-called critical Higgs inflation  ($\xi<100$) could be naturally realized 
and the SM vacuum stability problem can be solved. 
In order to achieve the gauge coupling unification around the string scale, 
we introduce new particles at TeV and intermediate scales.
Also, we employed the
Type I seesaw mechanism explaining the tiny neutrino masses. We have
shown that, by choosing the Seesaw scale, we can control
the SM Higgs quartic coupling at the inflation scale. 

We present a few benchmark points where we show that the scalar spectral indices
are around 0.9626 and 0.9685 for the number of e-folding
$N=50$ and $N=60$ respectively. The tensor-to-scalar ratios are order
of $10^{-3}$. The running of the scalar spectral index is negative
and is order of $10^{-4}$.

\begin{acknowledgments}
This research was supported in part by the Projects 11475238, 11647601 and 11605049 supported by National Natural Science Foundation of China,
and by Key Research Program of Frontier Science, CAS. The numerical
results described in this paper have been obtained via the HPC Cluster
of ITP-CAS. The work of IG was supported in part by Bartol Research
Institute.
\end{acknowledgments}

\appendix

\section{The RGEs in the SM}
For three standard model (SM) gauge couplings, we employ the two-loop
RGEs~\cite{AMALDO1992374,Barger:1992ac}
\begin{eqnarray}
\dfrac{\dd g_{i}}{\dd\ln\mu} & = & \dfrac{g_{i}}{16\pi^{2}}\left[b_{i}^{SM}g_{i}^{2}+\dfrac{1}{16\pi^{2}}\left(\sum_{j=1}^{3}b_{ij}^{SM}g_{i}^{2}g_{j}^{2}-C_{i}^{t}g_{i}^{2}y_{t}^{2}\right)\right],\label{1}
\end{eqnarray}
where $g_{i}~(i=1,2,3)$ are the SM gauge couplings, $y_{t}$ is the
top quark Yukawa coupling, and
\begin{eqnarray}
b_{i}^{SM}=\left(\frac{41}{9},-\frac{19}{6},-7\right),~b_{ij}^{SM}=\begin{pmatrix}\frac{199}{50} & \frac{27}{10} & \frac{44}{5}\\
\frac{9}{10} & \frac{35}{6} & 12\\
\frac{11}{10} & \frac{9}{2} & -26
\end{pmatrix},~C_{i}^{t}=\left(\frac{17}{10},\frac{3}{2},2\right).
\end{eqnarray}
The RGE for the top quark Yukawa coupling is
\begin{eqnarray}
\dfrac{\dd y_{t}}{\dd\ln\mu} & = & y_{t}\left(\frac{1}{16\pi^{2}}\beta_{t}^{(1)}+\dfrac{1}{(16\pi^{2})^{2}}\beta_{t}^{(2)}\right)\label{3}
\end{eqnarray}
with the one-loop and two-loop contributions given by
\begin{eqnarray}
\beta_{t}^{(1)} & = & -\sum c_{i}^{SM}g_{i}^{2}+\frac{3}{2}y_{t}^{2}+Y_{2},\\
\beta_{t}^{(2)} & = & \frac{1187g_{1}^{4}}{600}-\frac{23g_{2}^{4}}{4}-108g_{3}^{4}-\frac{3}{20}g_{2}^{2}g_{1}^{2}+\frac{19}{15}g_{3}^{2}g_{1}^{2}+9g_{2}^{2}g_{3}^{2}\nonumber \\
 &  & +y_{t}^{2}\left(\frac{223}{80}g_{1}^{2}+\frac{135}{16}g_{2}^{2}+16g_{3}^{2}\right)\nonumber \\
 &  & +\frac{5}{2}Y_{4}-6\lambda y_{t}^{2}+\frac{3}{2}y_{t}^{4}-\frac{9}{4}Y_{2}y_{t}^{2}-\chi_{4}+\frac{3}{2}\lambda^{2},
\end{eqnarray}
where $Y_{2}=3y_{t}^{2}$, $Y_{4}=\sum_{i=1}^{3}c_{i}^{SM}g_{i}^{2}y_{t}^{2}$,
$\chi=\frac{27}{4}y_{t}^{4}$, and $c_{i}^{SM}=\left(\frac{17}{20},\frac{9}{4},8\right)$.
The RGE for the Higgs boson quartic coupling is
\begin{eqnarray}
\dfrac{\dd\lambda}{\dd\ln\mu} & = & \frac{1}{16\pi^{2}}\beta_{\lambda}^{(1)}+\dfrac{1}{(16\pi^{2})^{2}}\beta_{\lambda}^{(2)},\label{A6}
\end{eqnarray}
where the one-loop and two-loop contributions are
\begin{eqnarray}
\beta_{\lambda}^{(1)} & = & 12\lambda^{2}+12y_{t}^{2}\lambda-12y_{t}^{4}-\left(\frac{9}{5}g_{1}^{2}+9g_{2}^{2}\right)\lambda+\nonumber \\
 &  & \frac{9}{4}\left(\frac{3}{25}g_{1}^{4}+\frac{2}{5}g_{1}^{2}g_{2}^{2}+g_{2}^{4}\right)\label{A7}\\
\beta_{\lambda}^{(2)} & = & -78\lambda^{3}+18\left(\frac{3}{5}g_{1}^{2}+3g_{2}^{2}\right)\lambda^{2}-\left(\frac{73}{8}g_{2}^{4}-\frac{117}{20}g_{1}^{2}g_{2}^{2}-\frac{1887}{200}g_{1}^{4}\right)\lambda-3\lambda y_{t}^{4}\nonumber \\
 &  & +\frac{305}{8}g_{2}^{6}-\frac{867}{120}g_{1}^{2}g_{2}^{4}-\frac{1677}{200}g_{1}^{4}g_{2}^{2}-\frac{3411}{1000}g_{1}^{6}-64g_{3}^{2}y_{t}^{4}-\frac{16}{5}g_{1}^{2}y_{t}^{4}-\frac{9}{2}g_{2}^{4}y_{t}^{2}\nonumber \\
 &  & +10\lambda y_{t}^{2}\left(\frac{17}{20}g_{1}^{2}+\frac{9}{4}g_{2}^{2}+8g_{3}^{2}\right)-\frac{3}{5}g_{1}^{2}y_{t}^{2}\left(\frac{57}{10}g_{1}^{2}-21g_{2}^{2}\right)\nonumber \\
 &  & -72\lambda^{2}y_{t}^{2}+60y_{t}^{6}~.~\,
\end{eqnarray}

\section{Additional contribution to the SM RGEs}

The one-loop contributions to the beta function coefficients from
the new particles are ~\cite{GCU,Barger:2007qb}
\begin{eqnarray*}
 &  & \Delta b^{Q_{x}+Q_{x}^{c}}=\left(\frac{2}{15},2,\frac{4}{3}\right)~,~~~\Delta b^{D_{x}+D_{x}^{c}}=\left(\frac{4}{15},0,\frac{2}{3}\right)~,~\\
 &  & \Delta b^{G_{x}}=(0,0,2)~,~~~\Delta b^{W_{x}}=\left(0,\frac{4}{3},0\right)~,~
\end{eqnarray*}
and the two-loop contributions are
\begin{eqnarray}
\Delta B^{Q_{x}+Q_{x}^{c}}=\left(\begin{array}{ccc}
\frac{1}{150} & \frac{3}{10} & \frac{8}{15}\\
\frac{1}{10} & \frac{49}{2} & 8\\
\frac{1}{15} & 3 & \frac{76}{3}
\end{array}\right) & , & \Delta B^{D_{x}+D_{x}^{c}}=\left(\begin{array}{ccc}
\frac{4}{75} & 0 & \frac{16}{15}\\
0 & 0 & 0\\
\frac{2}{15} & 0 & \frac{38}{3}
\end{array}\right),\label{2a}\\
\Delta B^{G_{x}}=\left(\begin{array}{ccc}
0 & 0 & 0\\
0 & 0 & 0\\
0 & 0 & 48
\end{array}\right) & , & \Delta B^{W_{x}}=\left(\begin{array}{ccc}
0 & 0 & 0\\
0 & \frac{64}{3} & 0\\
0 & 0 & 0
\end{array}\right).
\end{eqnarray}
The mass scales of the vector-like fermions $(Q_{x},~Q_{x}^{c})$
and $(D_{x},~D_{x}^{c})$ are set to be $M_{V}=1~{\rm TeV}$, and
we denote the masses of $G_{x}$ and $W_{x}$ as $M_{8}$ and $M_{3}$,
respectively, which will be chosen at the intermediate scales. When
renormalization scale $\mu$ is larger than $M_{V}$, $M_{3}$, or
$M_{8}$, the contributions of the corresponding particles should
be taken into account.

\section{Type I Seesaw Mechanism}

Here we present additional one-loop contributions to the various beta
function coefficients having Type I seesaw mechanism for neutrino
masses. We define
\[
S_{\nu}=Y_{\nu}^{\dagger}Y_{\nu}=Y_{\nu}^{T}Y_{\nu}=-\dfrac{2M_{R}}{v^{2}}M_{\nu},
\]
then additional one loop contribution to the top Yukawa and Higgs
quartic couplings are the following
\begin{eqnarray}
 &  & \delta\beta_{t}^{(1)}=\tr[S_{\nu}],\\
 &  & \delta\beta_{\lambda}^{(1)}=4\tr[S_{\nu}]\lambda-4\tr[S_{\nu}^{2}]~.~\,
\end{eqnarray}
Above the scale of right handed neutrino we have the following one
loop RGE for $S_{\nu}$
\begin{eqnarray}
\dfrac{\dd S_{\nu}}{\dd\ln\mu}=\dfrac{S_{\nu}}{16\pi^{2}}\left[6y_{t}^{2}+2\tr[S_{\nu}]-\left(\frac{9}{10}g_{1}^{2}+\frac{9}{2}g_{2}^{2}\right)+3S_{\nu}\right]\;.
\end{eqnarray}

\section{Phenomenological constraints and scanning procedure}

With $M_{V}$, $M_{3}$, $M_{8}$ and $M_{R}$ fixed, we solve the
RGEs to get the evolution of the SM Higgs quartic coupling from $M_{Z}$
to $M_{{\rm String}}=3\times10^{17}~$GeV or $M_{{\rm Pl}}=2.43\times10^{18}~{\rm GeV}$
which is the unification scale in our models. We integrate the SM
gauge coupling RGEs first in Eq.~(\ref{1}) with $y_{t}$ from $M_{Z}$
to $m_{t}$ to determine the initial value of the top quark Yukawa
coupling $y_{t}(m_{t})$. The running top quark mass $m_{t}$ gets
one-loop and two-loop QCD corrections and also the one-loop electroweak
one-loop correction. To solve the RGEs, we use the boundary conditions
at $M_{Z}$ given by
\begin{equation}
g_{1}(M_{Z})=\sqrt{\frac{5}{3}}\dfrac{g_{{\rm em}}}{\cos\theta_{W}}~,\;g(M_{Z})=\dfrac{g_{{\rm em}}}{\sin\theta_{W}}~,\;g(M_{Z})=\sqrt{4\pi\alpha_{3}}\;,\;y_{t}(m_{t})=\dfrac{\sqrt{2}M_{t}(m_{t})}{v},\label{dx}
\end{equation}

We then integrate the RGEs for $(g_{i},y_{t},\lambda)$ from $M_{Z}$
to the scales of vector-like fermions $M_{V}$, $M_{3}$, $M_{8}$
and to the Seesaw scale $M_{R}$ and continue to integrate the RGEs
for $(g_{i},y_{t},\lambda,S_{\nu})$ from $M_{R}$ to $M_{String}/M_{Pl}$.


\end{document}